\documentclass[aps,pra,twocolumn,superscriptaddress,nofootinbibt,longbibliography]{revtex4-2}
\usepackage[colorlinks=true, citecolor=red, urlcolor=blue ]{hyperref}
\usepackage[T1]{fontenc}
\usepackage{amsmath,amssymb}
\usepackage{amsfonts}
\usepackage{epsfig,pstricks,graphicx}
\usepackage{bm}
\usepackage{physics}
\usepackage{array}
\usepackage{diagbox}
\usepackage{multirow}
\usepackage{hyperref}

\usepackage{color}
\usepackage{array}
\usepackage{bbm}
\usepackage{ulem}

	\definecolor{indiagreen}{rgb}{0.07, 0.53, 0.03}
	\definecolor{x}{rgb}{0.8, 0.0, 0.7}
\begin{document}

\title{Optimal interferometry for Bell-nonclassicality induced by a vacuum-one-photon qubit}
 
\author{Tamoghna Das}
\author{Marcin Karczewski}
\author{Antonio Mandarino}
\author{Marcin Markiewicz}
\author{Marek \.Zukowski}
\affiliation{International Centre for Theory of Quantum Technologies, University of Gda\'nsk, 80-308 Gda\'nsk, Poland}

\begin{abstract}
Bell nonclassicality of a single photon superposition in two modes, often referred to as `nonlocality of a single photon', is one of the most striking nonclassical phenomena discussed in the context of foundations of quantum physics. Here we show how to robustly violate local realism within the weak-field homodyne measurement scheme for \textit{any} superposition of one photon with vacuum. Our modification of the previously proposed
setups involves tunable beamsplitters at the measurement stations, and the local oscillator fields significantly varying between the settings, optimally being {\it on} or {\it off}. As photon number resolving measurements are now feasible, we advocate for the use of the Clauser-Horne Bell inequalities for detection events using precisely defined numbers of photons. We find a condition for optimal measurement settings for the maximal violation of the Clauser-Horne inequality with weak-field homodyne detection, which states that the reflectivity of the local beamsplitter must be equal to the strength of the local oscillator field. We show that this condition holds not only for the vacuum-one-photon qubit input state, but also for the  superposition of a photon pair with vacuum, which suggests its generality as a property of weak-field homodyne detection with photon-number resolution.
Our findings suggest a possible path to employ such scenarios in device-independent quantum protocols.

\end{abstract}

\maketitle

\section{Introduction}

Security of many quantum information protocols relies on a violation of local realism (often imprecisely referred to as `nonlocality'). In the standard approach, it results from the incompatibility of measurements and the entanglement between at least two particles \cite{HHH09,ZukowskiRMP,Brunner14}. However, the idea of violating local realism with just a single particle has also been extensively investigated \cite{TWC91, Hardy94, Vaidman95, Hessmo04, Enk05, Dunningham07, Heaney11, Jones11, Brask13, Morin13, Fuwa2015, Lee17,1stPaper, 2ndPaper}.

First experimental proposals aimed at demonstrating
the ``nonlocality of a single photon''  \cite{oliver1989, TWC91} employed weak balanced homodyne measurements, with local oscillators of fixed strength whose phases defined the local settings. After a long debate, these schemes were recently decisively rejected, as a local hidden variable model replicating \textit{all} their outcome probabilities exists \cite{1stPaper, 2ndPaper}. 

\begin{figure} \includegraphics[width= 1 \columnwidth]{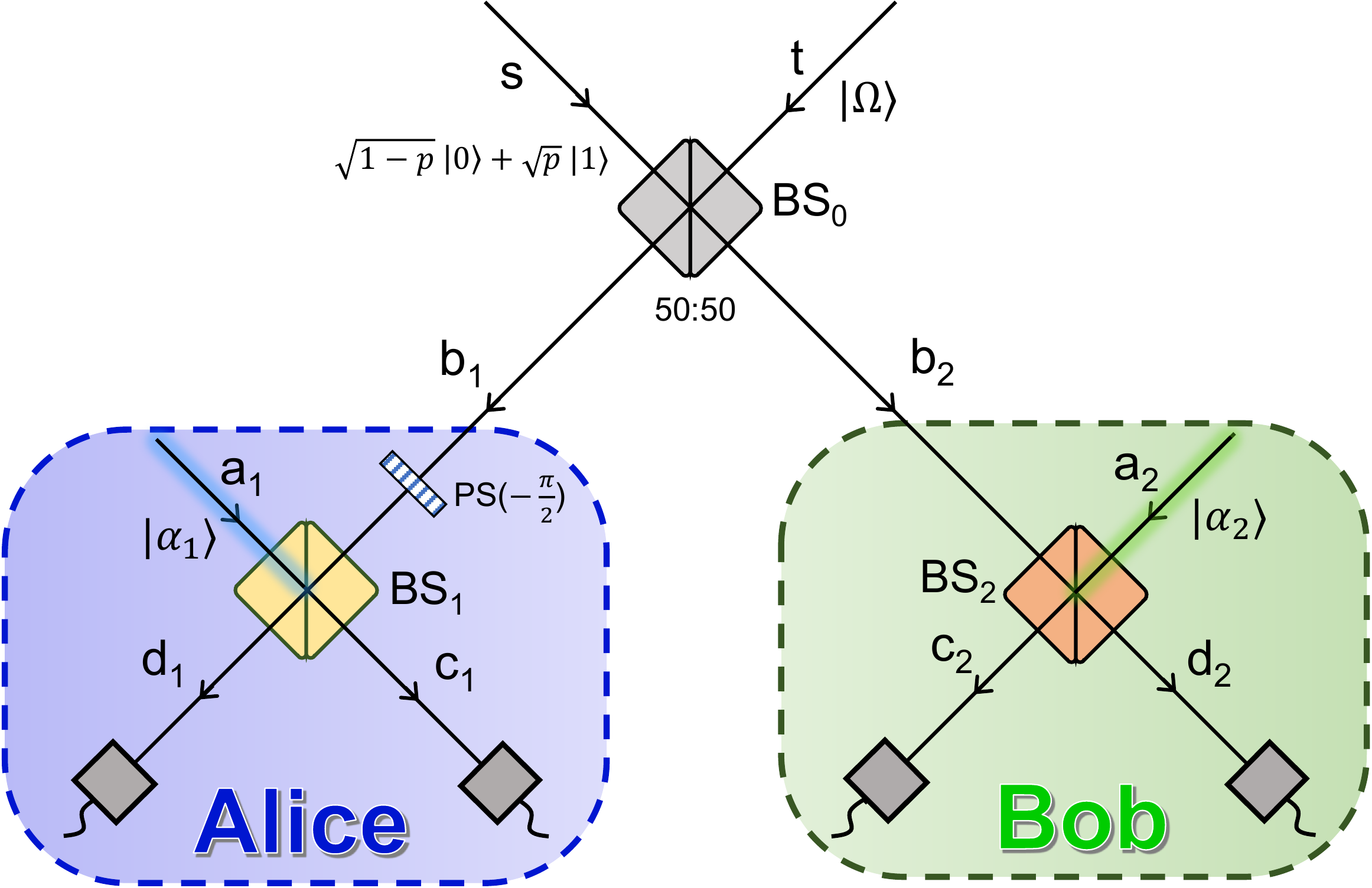}
	\caption{\label{mainSetup}
			Schematic representation of the general homodyne setup for testing violation of local realism by the vacuum-one-photon state. The original proposal of Hardy results from fixing $BS_i$ ($i=1,2$) as balanced beamsplitters for one measurement setting and removing them for the other. PS is a phase shifter that compensates the $\pi/2$ phase acquired by the reflected beam after BS$_0$.}
\end{figure}

A modification of these schemes, put forward by Hardy in \cite{Hardy94},
showed genuine violation of local realism by the state
\begin{equation}
\label{eq:InState}
  \ket{\psi(p)} = \sqrt{1-p}\ket{00}_{b_1,b_2} + \sqrt{\frac{p}{2}} \left(\ket{01}_{b_1,b_2} + \ket{10}_{b_1,b_2} \right),
\end{equation}
for  $0<p\leq1$, presented here without irrelevant phase factors.
In the formula, e.g. $\ket{01}_{b_1, b_2}$ denotes single excitation of the field mode $b_2$ and no excitation of the mode $b_1$.
This state is obtained by feeding one input of a balanced beamsplitter with a vacuum-one-photon qubit (the state given by the superposition of the vacuum and a single photon) and the other one with the vacuum, as presented in Fig. \ref{mainSetup}.
Aside from the initial state, Hardy's scheme differs from the ones of \cite{oliver1989, TWC91} in turning local oscillators \textit{on} or \textit{off}, depending on the measurement settings.

The violation in \cite{Hardy94} is presented in the form of a paradox and relies on precisely tuned local oscillators that interfere destructively with the initial state, completely erasing specific terms appearing after $BS_1$ and $BS_2$ (see Fig. \ref{mainSetup}).
Such reasoning, now customarily called `Hardy's paradox', is very elegant but does not necessarily lead to the most robust violations of Bell inequalities.

This begs the question of whether further modifications of the setup proposed by Hardy could improve the predicted Bell violations for $\ket{\psi(p)}$, especially for $p=1$, where the original scheme falls short. 
In this work, we investigate all-optical schemes involving only local oscillators, tunable beamsplitters, and photon number resolving detection. A  modification of the original Hardy's setup allows us to violate  a Clauser-Horne (CH) Bell inequality 
    for all non-zero values of $p$. We also show that the original Hardy's approach works {\it effectively} only for non-trivial superpositions of a single-photon state, i.e. 
    for a non-negligible  amplitude of the vacuum component. If its probabilistic weight is  around $8\%$ or less, the Hardy approach leads to experimentally irrelevant, minute violations of local realism. We show that tunable beamsplitters are  essential to violate local realism robustly, and the local oscillator fields must significantly vary between the settings. The optimal arrangement is such that the local oscillators are switched  \textit{off} in the case of one of the local measurement settings. 
 As photon number resolving measurements are now feasible, see, e.g. \cite{Walmsley14, Walmsley20}, we use Clauser-Horne Bell inequalities for detection events using precisely defined numbers of photons. This sets a possible path to apply such scenarios in device-independent quantum protocols. Finally, we obtain an interesting interferometric condition relating the power of the local oscillator and reflectivity of the local beamsplitter (they must be equal)
which allows the most robust violation of local realism for each value of $p$. 

Moreover, the single photon delocalized into two modes (the $p=1$ case in formula (\ref{eq:InState})) has been 
used to build 
reliable quantum repeaters networks \cite{Repeater1, Repeater2} 
that require less resources and are less sensitive to source of experimental inefficiencies than other platforms \cite{REV-Repeater}. Therefore
we believe that our work constitutes a basic ingredient to devise quantum information protocols based on Bell-nonclassicality with minimal physical resources, such as device-independent quantum key distribution or self-testing. 

\section{Preliminaries}

\subsection{Experimental setup} \label{exp_set} 
We consider here an experimental scheme generalizing the setups of \cite{Hardy94, TWC91} by introducing tunable beamsplitters at the detection stations. This modification allows us to recover the original proposals as limiting cases, but also to go beyond them and determine a procedure that optimally detects Bell nonclassicality.    
The final measurements will be maintained all-optical, namely we use only passive optical elements and coherent local beams as auxiliary systems to implement the photon number resolving weak-field homodyne detection scheme.  
The  scheme uses three  beamsplitters BS$_j$ for $j=0,1,2$, see Fig. \ref{mainSetup}.
The first beamsplitter, BS$_0$, is a 50-50 one that serves for the preparation of the input state. The remaining two, BS$_1$ and BS$_2$, are tunable beamsplitters used by two spatially separated parties, Alice and Bob, in their local homodyne photon-number resolving measurements. Note that in the proposals of \cite{TWC91} and \cite{Hardy94} the beamsplitters were symmetric 50-50 ones. Such were used in experiment \cite{Walmsley14}.

{\bf State preparation --} Consider two input modes $s$ and $t$ of  BS$_0$. The input mode $s$ is
{in} a vacuum--single-photon qubit  {state}, namely it is a superposition  
\begin{equation}
\label{eq:input}
    \sqrt{1-p} |0\rangle_s + \sqrt{p}|1\rangle_s,
\end{equation} 
where $0 < p \leq 1$. The  input mode $t$ is  in the vacuum state. 
After passing through BS$_0$ the output state in modes $b_1$ and $b_2$ is given by (\ref{eq:InState}), which can also be written in an equivalent  form of:
\begin{equation}
\label{HardyPsi}
 \ket{\psi(p)} = \left( \sqrt{1-p} + \sqrt{\frac{p}{2} } \left(\hat b_1^\dagger + \hat b_2^\dagger \right)\right)|\Omega\rangle,
\end{equation}
where $\hat b_j^\dagger$ is the creation operator in mode $b_j$, for $j = 1,2$ and $|\Omega\rangle$ is the vacuum state. 
For the sake of having a symmetric initial state, we choose  a mode transformation by the beamsplitter $BS_0$ of to lead to no relative phase shifts between reflected and transmitted beams in the case of radiation entering via input $s$. The case presented in the figure 1 shows a symmetric 50-50 beamsplitter, whose phase jump $\pi/2$ at the reflection is compensated by a suitable phase shifter, PS.
The case of an unbalanced BS$_0$ will be discussed in Section \ref{sec:generalized}.

{\bf Measurement stations --}
The optical field in state (\ref{HardyPsi}) is sent to two spacelike-separated observers Alice and Bob. 
The local measurement station use tunable beamsplitters $BS_1$ and $BS_2$,
and auxiliary weak coherent local oscillator fields $\ket{\alpha_j}_{a_j}$  with amplitudes $
\alpha_j$ and phases $\phi_j$, which are fed into  the input ports $a_j$,  where $j =1,2$. The moduli of amplitudes of the local oscillators reaching the beamsplitters are assumed to be tunable, as in the case of \cite{Hardy94}. In Ref. \cite{TWC91} they were constant.

In the proposals of Refs. \cite{TWC91} and \cite{Hardy94} the final beamsplitters were 50-50 ones. Here, we  consider tunable two-input-two-ouput beamsplitters. They can be realized with e.g. Mach-Zehnder interferometers.

We  assume that the tunable beamspliters  perform  unitary transformations $U_{BSj}$ on the input  modes, given by:
\begin{equation}
\begin{pmatrix} \label{SU2trans2}
\hat  c_j \\
\hat  d_j 
\end{pmatrix}    
= U_{BS_j}(\chi_j)
\begin{pmatrix}
\hat a_j \\
\hat b_j 
\end{pmatrix}, 
\end{equation} 
with the unitary matrix:

\begin{equation} \label{SU2Un}
U_{BS}(\chi) = 
\begin{pmatrix}
\cos \chi & i\sin \chi \\
 i\sin \chi & \cos \chi
\end{pmatrix},   
\end{equation}
where $\cos  \chi = \sqrt{T}$ is the transmission amplitude of the beamsplitter, 
see e.g. \cite{bachor2019guide, Pan2012}. Then, detectors in modes $c_j$ and $d_j$ register numbers of photons. 

In the sequel special local settings will play a crucial role. 
Those with $T_j=1$ and $\alpha_j=0$, will be called the \textit{off} settings. Note that they are equivalent to no beamsplitter and  the local oscillator switched off. The remaining settings are called \textit{on}.  

\subsection{Clauser-Horne inequality  based on single photon detection}
\label{detection}

The Clauser-Horne \cite{CH74} inequality reads:
\begin{eqnarray}         \label{CHin}
   -1~\leq~P(A,B)+P(A,B')+P(A',B) -P(A',B') \nonumber \\
        -P(A)-P(B)~=~CH~\leq~0, ~~~
   \end{eqnarray}
where $A$ and $A'$ for Alice, $B$ and $B'$ for Bob denote some local events for different local settings, whereas $P(\cdot)$ and $P(\cdot,\cdot)$ denote a local probability and a joint probability respectively.

We propose, inspired by \cite{Hardy94}, that Alice's detection events $A$ and $A'$ are in both cases a single photon in mode $d_1$ and no photon in mode $c_1$ for her two possible measurement choices. Bob's events $B$ and $B'$ are defined analogously. Note that such measurement results are linked with \textit{one-dimensional projectors} onto a specific Fock state. 

The measurement settings, ``primed'' and ``unprimed'', are given by the amplitude $\alpha_j^{(')}$ and phase $\phi_j^{(')}$ of the coherent local oscillator fields, and the reflectivity $R_j = \sin^2 \chi_j^{(')}$
of the local beamsplitter ($j = 1,2$ for Alice and Bob respectively).

\begin{figure}[t]
	\centering
\includegraphics[width= 1
\columnwidth]{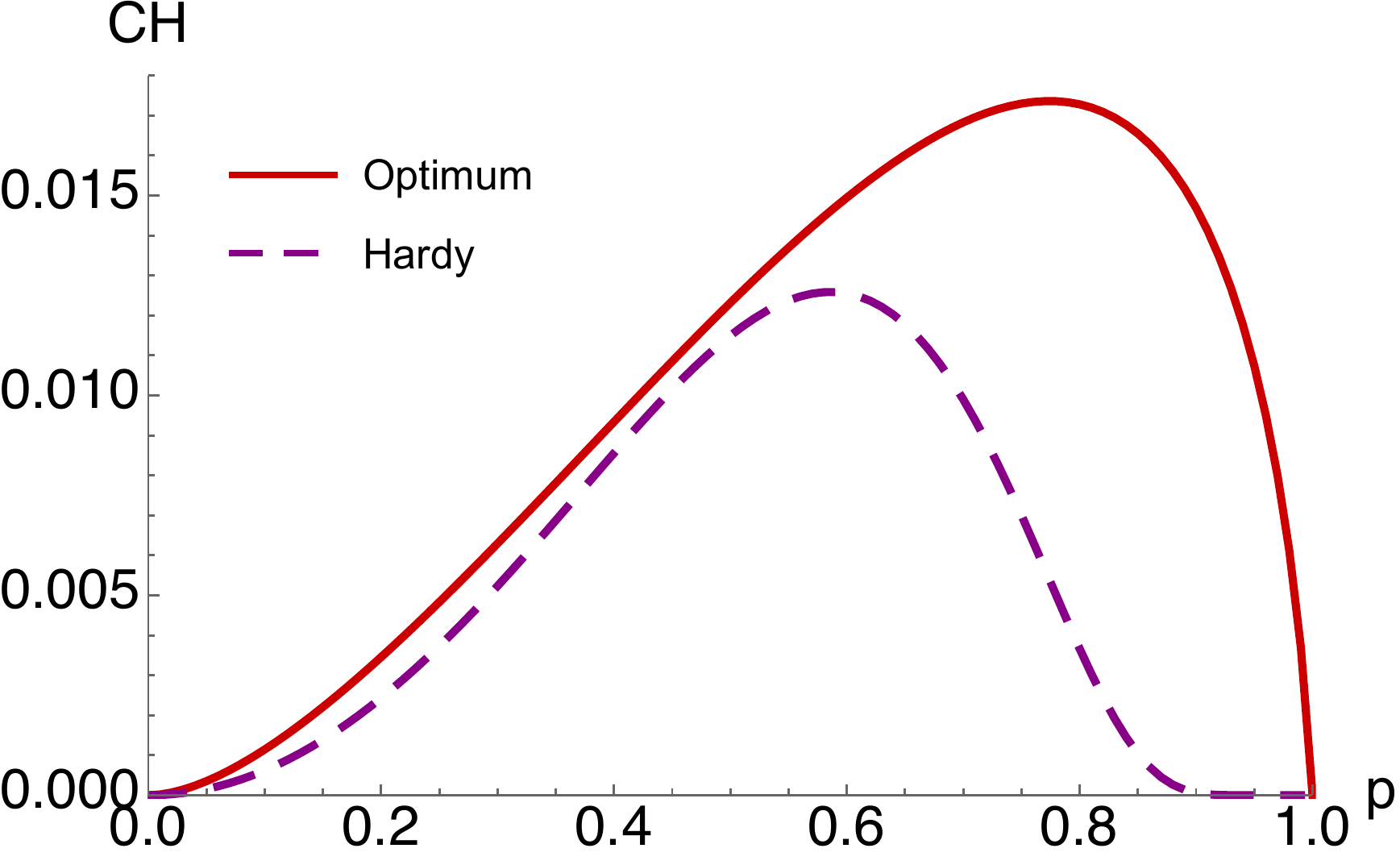}
	\caption{\label{CH_max}
		Maximal violation of the right hand side of the CH inequality as a function of the input state parameter $p$. The red solid curve corresponds to  the maximum for completely general measurement settings (any phases, any local oscillator amplitudes, any transmittivities of the final beamsplitter, for each possible local setting). The purple dashed curve represents the CH value for the Hardy's settings  of \cite{Hardy94}.}
\end{figure}

\section{Search for CH inequality violations: scan over arbitrary pairs of the settings}

We consider general measurement settings, of the above kind, for both Alice and Bob.  We do not fix any of the setting parameters: values of beamsplitter's reflectivity $R_j = \sin^2 \chi_j$ 
amplitudes of coherent states $\alpha_j$, and their phases $\phi_j$, neither for the unprimed setting nor for the primed one (on both sides).

The joint probability of detecting single photon in mode $d_j$ and no photon in mode $c_j$, for $j = 1,2$, for the unprimed settings, is given by:
\begin{eqnarray}
&&P(A,B) = \nonumber \\&& \hspace{-1em}|{}_{c_1,d_1,c_2,d_2}\bra{0,1,0,1}(U_{BS_1})_{a_1,b_1}(U_{BS_2})_{a_2,b_2}\ket{\Psi_{p}(\alpha_1,\alpha_2)}|^2, \nonumber \\
\end{eqnarray}
where $\ket{\Psi_{p}(\alpha_1,\alpha_2)}$ is the full input state (\ref{eq:InState}), augmented with the auxiliary coherent fields. It is given by:
\begin{eqnarray}
    && \ket{\Psi_{p}(\alpha_1,\alpha_2)} \nonumber \\
   && \hspace{-0.5em} =  \ket{\alpha_1}_{a_1}\left(\sqrt{1-p}\ket{00} + \sqrt{\frac p2} (\ket{01} + \ket{10} )\right)_{b_1,b_2} \ket{\alpha_2}_{a_2}. ~~~~
\end{eqnarray}
The joint probability reads:
\begin{eqnarray}
\label{Pab}
 && P(A,B) =   e^{-\alpha_1^2 -\alpha_2^2}\Big[ (1 -p) R_1 R_2 \alpha_1^2 \alpha_2^2 \nonumber \\ 
 && + \frac{p}{2}\Big( \alpha_1^2 R_1 (1 - R_2) + (1 - R_1) \alpha_2^2 R_2  \nonumber \\ 
&&    + 2 \alpha_1 \alpha_2 \sqrt{R_1R_2(1 - R_1)  (1 - R_2)} \cos(\phi_1 - \phi_2) \Big)   \nonumber \\
&&  - \alpha_1 \alpha_2 \sqrt{2 p (1 - p) R_1R_2} \times \nonumber \\ 
&& \left(\sqrt{(1 - R_1)R_2} \alpha_2 \sin \phi_1 + \sqrt{R_1(1 - R_2)} \alpha_1 \sin \phi_2 \right)
\Big].~~~
\end{eqnarray}

We have used in the formula the reflectivities $R_j=\sin^2{\chi_j}$, as it turns out that in some calculations this is a better choice.
Probabilities  $P(A,B'), P(A',B)$ and $P(A',B')$ can be obtained from $P(A,B)$ by replacing the unprimed parameters by primed ones for this general set of measurements.

\begin{figure}[t]
	\centering
\includegraphics[width= 1
\columnwidth, height = 0.62 \columnwidth]{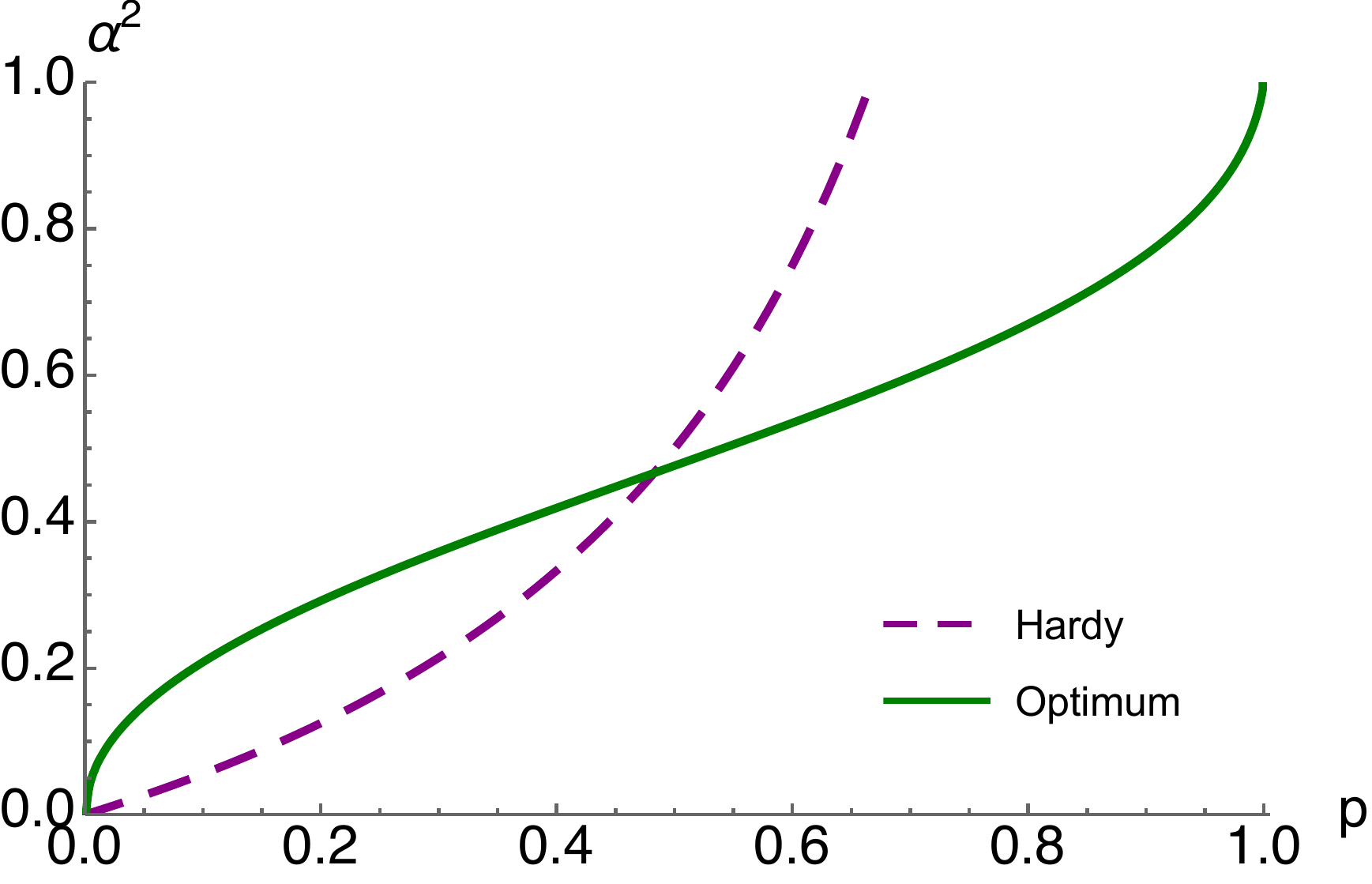}
	\caption{\label{Opt_alpha}
		Plot of the optimal coherent state strength $\alpha_0^2$, which maximizes the  CH expression given in (\ref{CHin}), for completely general measurement settings, as a function of $p$, depicted as a green line. 
		We compare our result with the coherent state strength  $\alpha^2 = \frac{p}{2(1-p)}$ (Hardy's measurement scheme), which is plotted by the dashed purple curve.
	 We have found that,  $\alpha^2$ of Hardy increases much faster than the optimal $\alpha_0^2$, and it is no longer in a  weak homodyne regime for $p \rightarrow 1$.  }
\end{figure}

The local probabilities are:
\begin{eqnarray}\label{Pa}
P(A) &=& \frac{1}{2} e^{-\alpha_1 ^2}\Big(p (1 - R_1) + \alpha_1 ^2 (2-p) R_1  \nonumber \\
&&  -  2\sqrt{2} \alpha_1  \sqrt{p(1-p)R_1(1-R_1)} \sin (\phi_1)  \Big),
\end{eqnarray}

\begin{eqnarray}
\label{Pb}
P(B) &=& \frac{1}{2} e^{-\alpha_2 ^2}\Big(p (1 - R_2) + \alpha_2 ^2 (2-p) R_2 \nonumber \\ 
&& - 2\sqrt{2} \alpha_2  \sin (\phi_2) \sqrt{p(1-p)R_2(1-R_2)} \sin (\phi_2) \Big). \nonumber \\
\end{eqnarray}

The optimization over all possible local parameters $\{\alpha_j, \phi_j, \chi_j\}$ for $j = 1,2$ and for primed and unprimed indices, yields $CH_{\max}$ as a function of $p$, which is depicted in Fig.  \ref{CH_max}. The measurement settings which give the maximum of the CH expression are the same for both $A$ and $B$ and for $A'$ and $B'$.
The primed settings  turn out to be the \textit{off} settings, with $T'_j = 1$, and $\alpha'_j = 0$, for both $j = 1,2$. Therefore the optimal settings indeed follow the \textit{on}/\textit{off} measurement scheme.
Note, that if primed events represent the {\it off} settings, we have 
\begin{eqnarray}
P(A,B') = \frac p2  R_1 \alpha_1^2 e^{-\alpha_1^2}, \\
P(A',B) = \frac p2 R_2 \alpha_2^2 e^{-\alpha_2^2},
\end{eqnarray}
and $P(A',B') = 0$. 

There is an interesting and puzzling  relation between the optimal transmissivity of the final beamsplitters and the intensity 
of the local coherent beams,
 which holds for the maximal violations.
Numerical results show that the optimal settings for CH inequality violation satisfy  $T_j + \alpha_j^2 = 1$, or equivalently 
\begin{equation}
R_j=\alpha_j^2, 
\label{OptCondR}
\end{equation}
for both $j = 1,2$, where $R_j=1-T_j$ is the reflectivity of the beamsplitter. This holds for  the entire range of $p$, except the $p \approx0$. However,  the $CH_{\max}$ in the region of $p \approx0$ is prohibitively small for the computer results to be reliable. One thus can conjecture that the condition for optimal settings $R_j=\alpha_j^2$ holds for all $p$. The discussion which follows strongly supports this conjecture. Also some analytical results, see further, lead to it. 

$R_j=\alpha_j^2$ seems to be a general optimality-of-settings pre-condition for ``single photon'' Bell experiments involving \textit{on}/\textit{off} weak homodyne measurements and single photon detection events. 
Although this relation does not look like a mere coincidence, we must admit that we were not able to find an underlying physical reason.
 As we shall see further, we have found this condition also to hold in the case of another experiment, which involves homodyne detection. 
 One additional comment is necessary here, namely we have fixed our single-photon detection events to one photon in mode $d_j$ and no photon in mode $c_j$, which is consistent with the original formulation of Hardy. However, we have verified that if we switch the modes in the above definition, namely we detect single photon in mode $c_j$ and no photons in mode $d_j$, the optimal CH values as a function of $p$ remain the same, whereas the optimality condition \eqref{OptCondR} changes into:
 \begin{equation}
     T_j=\alpha_j^2.
\label{OptCondT}
 \end{equation}
 This fact can be intuitively understood by noticing that for the switched single-photon detection scheme the condition \eqref{OptCondT} guarantees that the intensity of the auxiliary beam reaching the detector responsible for registering a single photon (here it is $c_j$) is the same as in the original scheme (for detector $d_j$ registering single photon). Therefore the two choices of single-photon detection scheme seem to be effectively symmetric. For clarity in the remaining part of the work we always keep the original convention for the detection events.

The optimal intensity of the coherent beam, denoted $\alpha_0^2$, is the same for both Alice and Bob. It  is plotted in Fig. \ref{Opt_alpha} by a solid green  curve as a function of $p$, which represents the fidelity of the input state (\ref{eq:input}) with the single-photon state.
A curve fitting on $\alpha_0^2$ yields 
an approximated functional dependence 
on $p$ given by:

\begin{equation}\label{eq:optalphamax}
    \alpha_0^2(p) \approx 1- \left(1-p^{0.39634}\right)^{0.453581}.
\end{equation}  

In our full unconstrained numerical analysis, some of our optimal parameters match with the not-optimized measurement settings in the Hardy's scheme \cite{Hardy94}. There, full  \textit{on}/\textit{off} settings are implemented using a balanced beamsplitter for both A and B (\textit{on} settings) or removing the beamsplitter and detecting one photon in mode $b_j, ~~j = 1,2$ (\textit{off} settings). 
A detailed analysis of Hardy's reasoning is given in Appendix \ref{s:hardyexp}. For a comparison we report here the value of  a violation of CH inequality related with the paradox of Hardy of Ref.  \cite{Hardy94}:
\begin{equation}\label{CH_Hardy}
   CH_{Hardy} =  \frac{e^{-\frac{p}{1-p}} p^2}{16 (1-p)},
 \end{equation}
which  is plotted in Fig. \ref{CH_max}, as a function of $p$, by dashed purple line. 
 The value of violation of the inequality for $p>0.922$ is prohibitively small, $CH_{Hardy} < 10^{-6}$, therefore it is of no significance in experiment, or for any 
 experimentally feasible protocol requiring a violation of local realism for its certification.
 Thus, the paradox of Hardy, from the point of view of experiment, pertains only to situations in which we have a significant 
 vacuum component in the signal beam $s$, Fig.1.

\begin{figure}[t]
	\centering
\includegraphics[width= 1
\columnwidth]{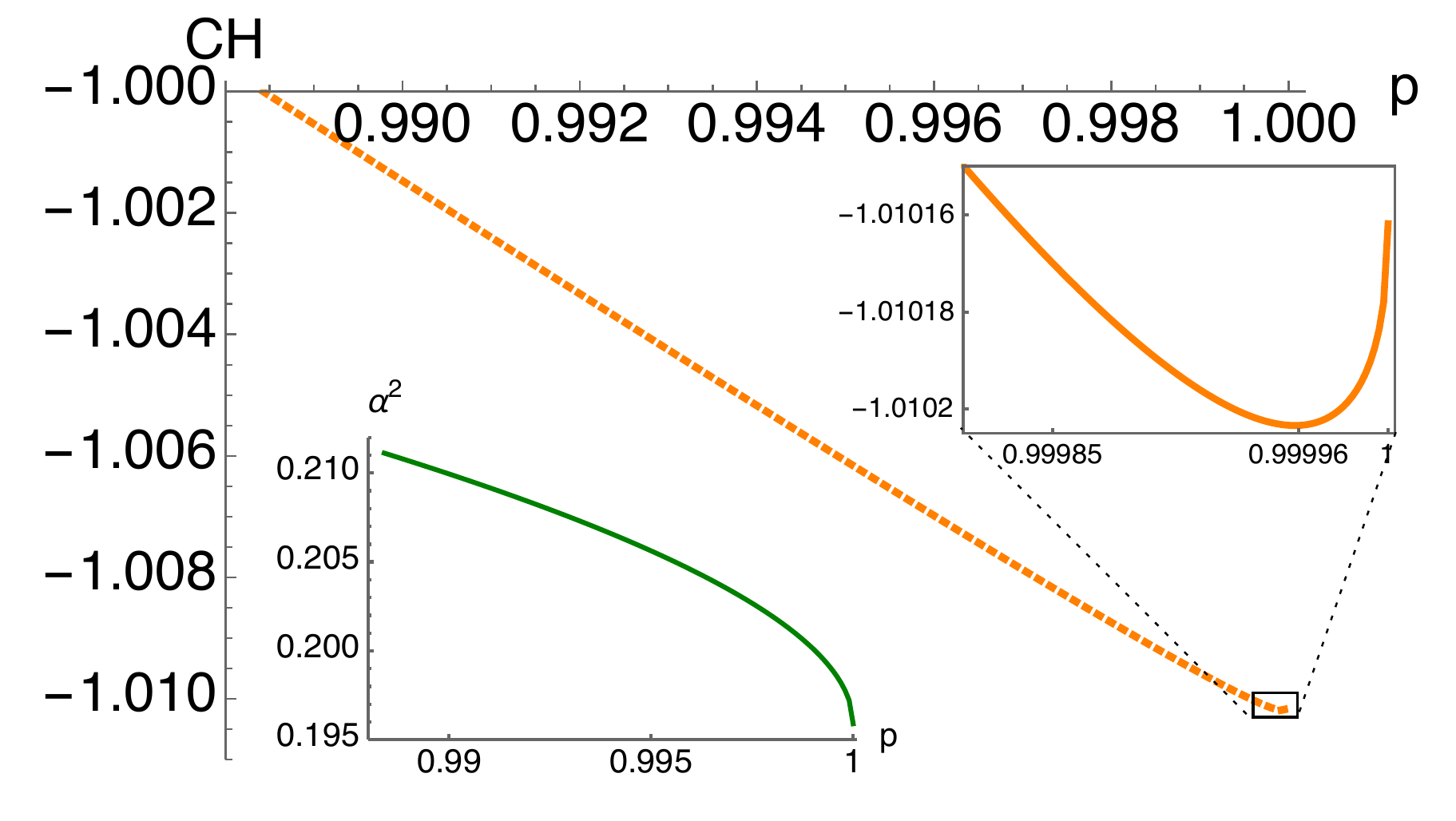}
	\caption{\label{CH_ours}
		Plot of the minimum $CH_{min}$ of the CH expression  as a function of $p$, which reveals the violation of the CH inequality in the negative side (orange dotted curve).  We obtained a measurable  violation of the CH inequality for the single photon limit for $ 0.989 \leq p \leq 1$, and hence, the Bell nonclassicality of (\ref{eq:InState}) can be experimentally detectable. The $CH_{min}$ is not a monotonic function of $p$, and it reaches its  minimum at $p = 0.999958$. The neighbourhood of it has been zoomed in. The variation of the optimal coherent state strength $\alpha_0^2$, has been plotted in the bottom left inset, which  always satisfy $ \alpha_0^2 = R_0$, $R_0$ being the optimal reflectivity of the local beamsplitter.}
\end{figure}

 Moreover, the optimal intensity of the local oscillator, which gives rise to $CH_{max}$, (solid green  curve in Fig. \ref{Opt_alpha}), lies entirely within the weak homodyne region. This is in contrast to the intensity of the coherent field used by Hardy \cite{Hardy94}, $\alpha_{Hardy}^2 = \frac{p}{2(1-p)}$, which goes to infinity as $p \rightarrow 1$. Hence, the investigation by Hardy for $p\approx 1$ is no longer in the regime of weak homodyning. \\

\subsection{No-vacuum and almost no-vacuum component input states ($p\approx1$)}\label{sec:peq1}

It turns out that the  CH inequality (\ref{CHin}) is non-negligibly violated on the left hand side for input states with small vacuum component, $0.989 < p \leq 1$.  This violation is most robust close to  $p=1$, however surprisingly not at $p=1$, where it is high but not maximal.
Its numerically obtained minimal values, $CH_{\min}$, are plotted in Fig. \ref{CH_ours}.

Due to the complexity of the CH expression (\ref{CHin}) it is hard to minimize it analytically. 
However, our numerical results point at several properties of the optimal settings.
Our unconstrained search for the minimal value of the CH expression always leads to the symmetric conditions:  
$\alpha_1=\alpha_2=0$ and $\chi_1=\chi_2=0$ (\textit{off} settings),  $\alpha'_1=\alpha'_2=\alpha$, $\chi'_1=\chi'_2 = \chi$, and $\phi'_1=\phi'_2=\frac{\pi}{2}$ (\textit{on} settings). 
Note that the optimal settings in this case also follow the \textit{on}/\textit{off} scheme, however now the primed settings are \textit{on}, whereas the unprimed -- \textit{off}.

Taking into account the above symmetry of the optimal settings for the single-photon input state ($p=1$), we get the following functional dependence of the value of the CH expression:
\begin{equation}
CH_{p=1} =-1+ \left[e^{-\alpha ^2} \alpha ^2  -2 e^{-2 \alpha ^2} \alpha ^2 (1-R)\right] R.
\end{equation}

Note that the $-1$ term is due to the fact that for the \textit{off} settings $P(A)=P(B)=\frac{1}{2}.$
In order to find its minimal value, we check the conditions under which the partial derivatives satisfy $\frac{\partial }{\partial R }CH_{p=1}=0$ and $\frac{\partial }{\partial \alpha }CH_{p=1}=0$.  The two  equations yield $R=\alpha^2$  as a necessary condition for the minimum. 
The threshold $\alpha$ which gives violation is given by equation: $e^{\alpha ^2} = 2(1 - 2 \alpha ^2)$, and the value of the violation is given by $[e^{-\alpha ^2} \alpha ^2  -2 e^{-2 \alpha ^2} \alpha ^2 (1-\alpha^2)] \alpha^2.$

The condition $R=\alpha^2$ is exactly the same relation between the transmissivity of a local beamsplitter and the local oscillator strength \eqref{OptCondR} that has been purely numerically found in the previous case of the right-side violation of the CH inequality. 
The latter one implies that in the optimal case  $\alpha ^2 \approx 0.1959$, which leads to  $CH_{min} \approx -1.01016$. 
Numerical calculations for $0.989 < p \leq 1$ also give $R=\alpha^2$ as the optimality condition.

\subsection{No violation of CH inequality based on fixed local  detection events of more than one photon for the {\it on}/{\it off} scenario} 

In the previous section, we have considered the violation of the CH inequality based on the coincidence detection of a single photon in mode $d_j$ and no photon in mode $c_j$, for both Alice and Bob, and for all local settings of the interferometric setup. 
In this section,  we will show that this is the only set of photon number detection events with which the Bell-nonclassicality of the single photon input state in (\ref{eq:InState}) can be revealed 
in a scenario with fixed photon-number detections for both local settings on both sides, and the \textit{on}/\textit{off} settings.

The local detection events will be denoted by $(n,m)$, where $n$ is the number of photons detected in the local detector $c_j$ and $m$ in $d_j$. Assume that all events $A$, $A'$, $B$, $B'$ are of this kind. We analyze below the cases  $(n,m)$ which are not $(0,1)$, $(1,0)$ (already analysed in previous section), or $(0,0)$ (which is trivial).

We consider three cases: 

\textbf{Settings $A, B$ are \textit{on} and $A', B'$ are \textit{off}:}
For the initial state (\ref{eq:InState}) and  local events $(n,m)$, with $n+m>1$, the following conditions trivially hold:
$$ P(A,B') = P(A',B) = P(A',B') = 0.$$ 
Thus, the CH expression now reduces to:  
\begin{equation}\label{reducedCH}
    CH = P(A,B) - P(A) - P(B),
\end{equation}
where the joint probability
\begin{eqnarray}
P(A,B) &=& P(n,m;n,m)_{c_1,d_1;c_2,d_2}, 
\end{eqnarray}
can be obtained from Eq. (\ref{Pab_n1m1n2m2}), of Appendix \ref{appen:arb_prob}, by putting $n_1 = n_2 = n$ and $m_1 = m_2 = m$.
And the local probabilities are 
\begin{eqnarray}
P(A) &=& P(n,m)_{c_1,d_1}, \\
P(B) &=& P(n,m)_{c_2,d_2},
\end{eqnarray}
see Appendix \ref{appen:arb_prob}  for the full expression.
The r.h.s of (\ref{reducedCH}) is always less than or equal to $0$ for any probabilistic theory. Moreover, the lower bound of the expression is for the studied problem always higher than  $-1$. 
 
This is because  the probability that Alice (or Bob) gets exactly $n+m>1$ particles on their side is less than $\frac{1}{2}$, and thus we have $P(A) + P(B) \leq 1$.

\textbf{Settings $A, B$ are \textit{off} and $A', B'$ are \textit{on}:}
In this case all probabilities involving $A$ or $B$ vanish, and the CH inequality reduces to
\begin{equation}
   -1 \leq CH = - P(A',B') \leq 0.
\end{equation}
Note that in this case the joint probability for the primed settings is $P(A',B') = P(n,m;n,m)_{c_1,d_1;c_2,d_2}$, (see Appendix \ref{appen:arb_prob})

\textbf{Settings $A, B$ and $A', B'$ are arbitrary:}
We have numerically checked that for $n+m > 1$, there is no violation of the CH inequality in the either side. This shows that condition $n+m=1$ gives the only set of events which reveals the nonclassicality of a single photon input state, if the detection scheme  events are fixed for both local settings and both observers.\\

\begin{figure}[h]
	\centering
\includegraphics[width= 1
\columnwidth]{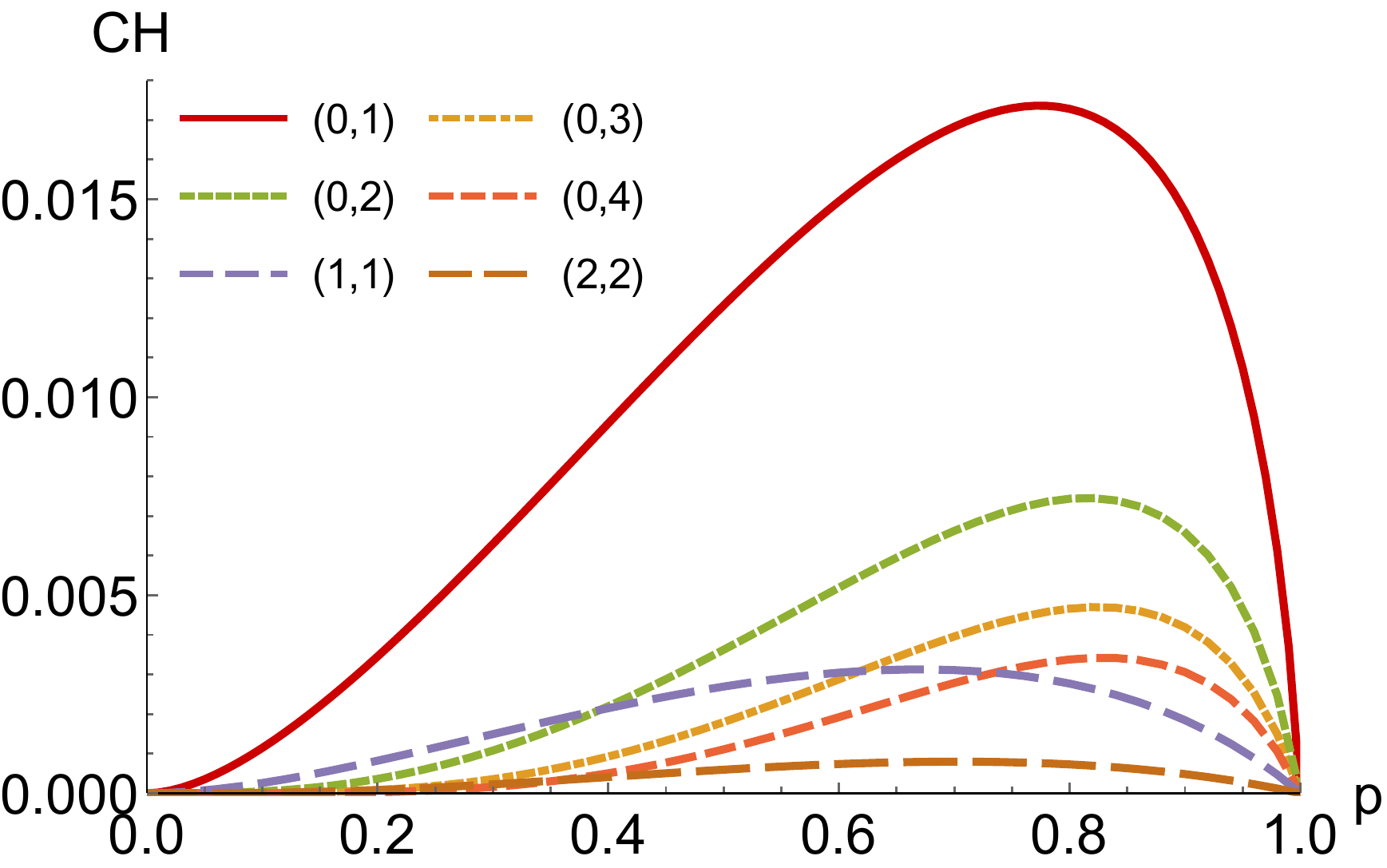}
	\caption{\label{CHallevents}
		 Plot of the maximal violation of the CH inequality for various values of $(n,m) \in \mathbb{N}^2$ representing detected local photon numbers in the \textit{on} setting, in the positive side of (\ref{CHin}). In this case we detect just a single photon in the \textit{off} case, as in the basic detection scenario we have proposed. The solid red line is the same as the one shown in Fig. \ref{CH_max}. 
		 }
\end{figure}

\subsection{CH inequality violation for different detection events associated with different settings, {\it on}/{\it off} scenario}

We will consider a CH inequality based on the \textit{on}/\textit{off} measurement settings scenario, however with differently defined detection events for the \textit{on} and \textit{off} case. For the \textit{on} settings we assume the pair of numbers $(n,m)$, with $m+n>1$ as the set of local detection events, in the same way as before, whereas for the \textit{off} case, i.e., when the beamsplitter is absent and the local oscillator is turned off,  the observer will detect only a single photon in mode $b_j=d_j$.

The joint probabilities for these new sets of events are as follows:
$P(A,B)$ is the same as in the previous case defined in Eq. (\ref{Pab}), 
$P(A,B')$ reads:
\begin{equation}
P(A,B')=\frac{p}{2 m! n!}e^{-\alpha_1^2}  \alpha_1^{2 (m+n)} R_1 {}^m (1 - R_1)^n,~~~
\end{equation}
and $P(A',B)$ is the same as $P(A,B')$ with $1 \leftrightarrow 2$ interchange. 

An optimization over all possible local parameters has been carried out for various values of $(n,m) \in \mathbb{N}^2$. We have observed that although there is a violation for nearly all possible values of $(n,m)$ \footnote{Like the previous section, there is no violation for the vacuum event, i.e., for $n = m = 0$.}, its magnitude is significantly smaller than for the $n+m=1$ case. The plot of the maximal violation of the CH inequality  (\ref{CHin}), for various values of $(n,m)$, is given in Fig. \ref{CHallevents}. It can be seen  that the maximal violation of the CH is decreasing with the increasing number of $m$ in the case of $n = 0$. This shows that for the single photon input state, detecting single photon in mode $d_j$ and no photon in mode $c_j$ (or \textit{vice versa}) is the optimal measurement choice to experimental detection of the vacuum-single-photon Bell nonclassicality.

Moreover, we found that in the case of the $(n,m)$ events for which $n+m>1$ the optimal $\alpha_0^2$ does not satisfy the condition $\alpha_0^2 = R$, and  $\alpha_0^2 > 1$ for a wide range of values of $p$.

The results of this section support our interpretation of the discussed experimental setup presented
in our previous work \cite{2ndPaper}, in which we interpret the nonclassicality found for the single-photon input state as arising from interference due to indistinguishability of photons from the input state and 
the local oscillator. Indeed, if we detect only a single photon at each local measurement station in the \textit{on}
setting, it must have come either from the input state or from the local oscillator. If we locally detect a higher number of photons, all but one of them must have come from the local oscillator. Thus, the indistinguishability of possible events is decreased, which suppresses the observed level of violation of local realism.
\begin{figure}[t]
	\centering
\includegraphics[width= 1
\columnwidth]{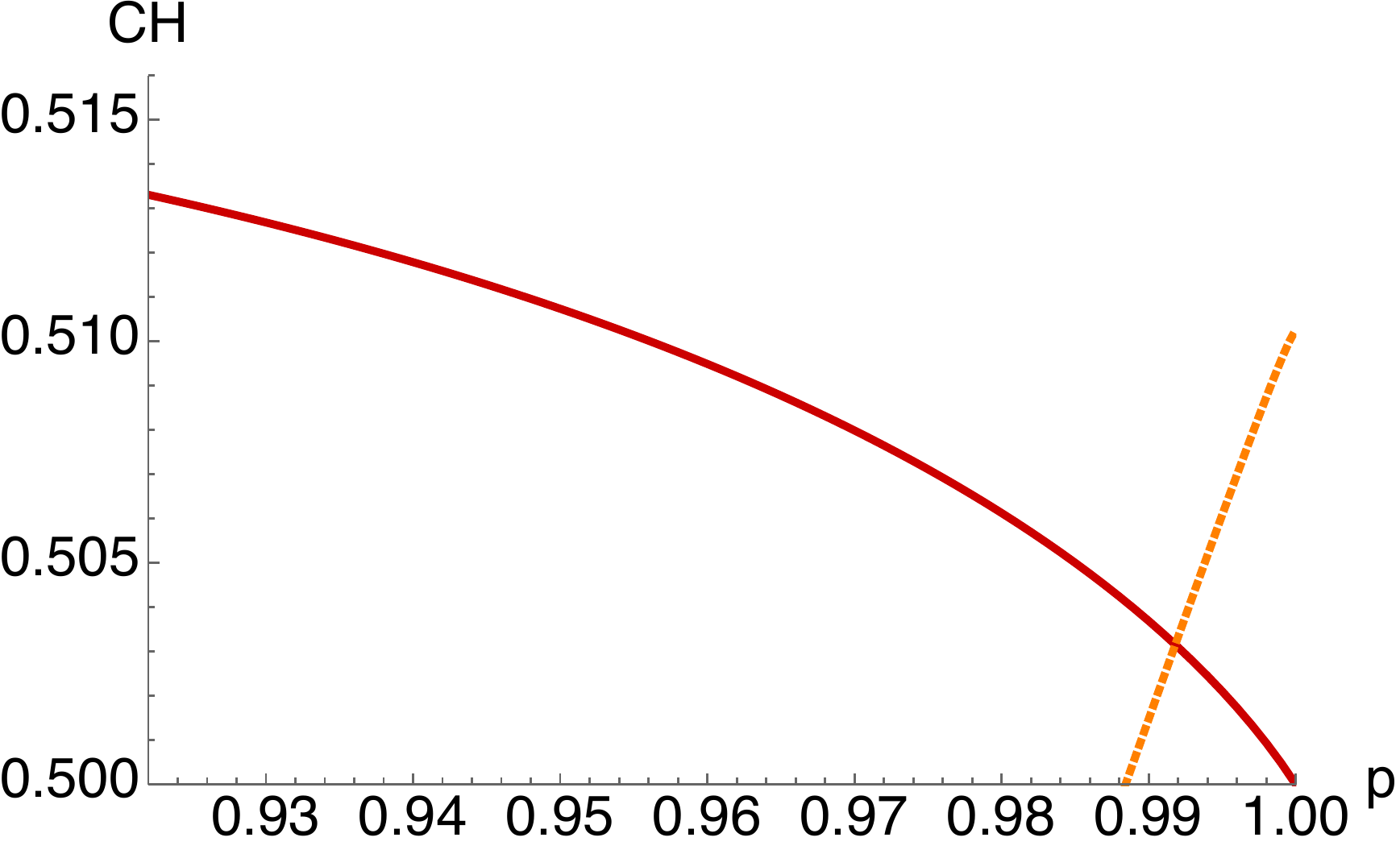}
	\caption{\label{CH_absolute}
		Plot of $\left|CH+\frac{1}{2}\right|$ as a function of $p$, for the one-parameter family of states (\ref{eq:InState}).
		 The range of $p$ shown in this plot is $0.922 \leq p \leq 1$, which is the region of no detectable violation by Hardy's scheme (see Fig. \ref{CH_max}). Here, the red solid curve is  the contribution of $CH_{\max}$, whereas the orange dotted line is coming from  $CH_{\min}$. Both the curves are above the $\frac{1}{2}$ limit, showing the violation of the CH inequality in both sides. It is clear that when there is no violation in the positive side,  one can interchange the measurement settings and obtain the violation in the other side. 
		 }
\end{figure}

\subsection{Absolute violation}
Note that the CH inequality can be put in a symmetric form
\begin{equation}
\left|CH+\frac{1}{2}\right|\leq \frac{1}{2}.
\end{equation}
A comparison of violations of the lower and upper bounds of the CH inequality in the range of $0.922 \leq p \leq 1$, where there is no significant violation by Hardy's scheme, is plotted in Fig. \ref{CH_absolute}.
The higher values for the juxtaposed curves of Fig. \ref{CH_absolute} show the magnitude of violation of this inequality.

\begin{figure}[t]
	\centering
\includegraphics[width= 1
\columnwidth]{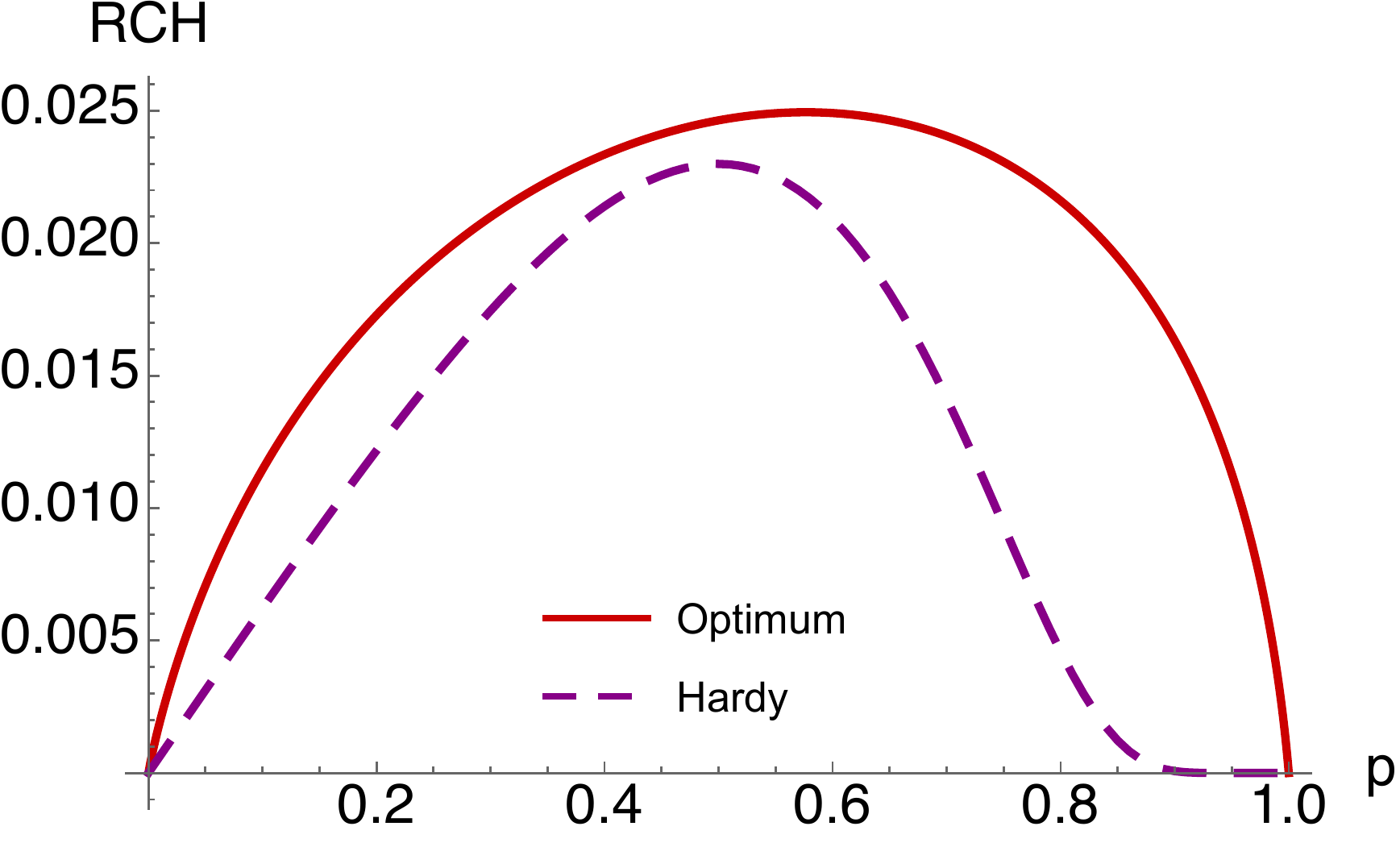}
	\caption{\label{RCH}
		Plot of the relative CH violation as a function of $p$. The term \textit{relative} has been used because we divide the maximal value of CH, given in (\ref{CHin}), by $p$, the probability of having single photon in the input state (\ref{eq:InState}). The red solid curve represent the optimum   relative CH violation, whereas the dashed purple curve is for the Hardy's scheme. }
\end{figure}

\subsection{Relative CH inequality for $p \approx 0$ }

The inequality we have derived in (\ref{CHin}) relies on a specific detection scheme that depends on the probability of detecting one photon in mode $d_j$ and no photon in mode $c_j $, for both $j = 1,2$. As it is easy to see in Fig. \ref{CH_max} the optimum CH value (for both our and Hardy's detection schemes)
signals that the violation is no longer significant for values $p \approx 0$.
Anyway, let us remind that $p$ quantifies the fidelity that the input state impinging on $BS_0$ has with the single photon states (see Eq. (\ref{eq:input})), so a low value of $p$ implies that the probability of detecting one photon in mode $d_j$ can be negligibly small and it affects the value of the violation of the CH inequality. To have a clear insight on the significance of the violation as function of $p$ along all its range, we plot the optimal CH value divided by the probability $p$ in in Fig. \ref{RCH} and term it as \textit{relative} CH value. This 
stress that our detection scheme still holds and allows to claim a violation of local realism also when the input state has a huge overlap with the vacuum.

\section{Robustness to experimental imperfections}

In the previous sections we provided a scheme to certify the ``non-locality'' of a single photon, or of a superposition of a single photon and vacuum. We found that when Clauser-Horne inequality is used, the "non-locality" is best detected by working with the \textit{on}/\textit{off} measurement settings and single photon detections. This extends the results of Hardy \cite{Hardy94} to a more general scenario. Whether a setting should be \textit{on} or \textit{off}  depends on the value of $p$, the probability of getting the single photon in the input state $|\psi(p)\rangle$. Moreover, we found that for the \textit{on} setting, the optimal transmittivities of the beamsplitters $BS_j$ are the same for both parties and they read $T_j = 1 - \alpha_0^2$. Here $\alpha_0^2$ denotes the optimal intensity of the local coherent state, which depends on $p$, as shown in Fig. \ref{Opt_alpha}, and is approximated in Eq. (\ref{eq:optalphamax}).

In this section we discuss the feasibility of implementing our scheme in inevitably imperfect experiments. We focus on two potentially most important sources of problems: fluctuation of the local fields around its optimal value and detector inefficiency. 

Note that, to achieve the optimal violation of the CH inequality, the parties need to tune their local settings: the reflectivity of beamsplitters, phases  of the auxiliary coherent fields and their intensities. The reflectivity is relatively easy to control and stable once set to the desired value. The outcome probabilities (\ref{Pab}-\ref{Pb}) depend on phases in a simple way. Thus, the effects of phase detuning will not differ from the ones seen in standard interferometric experiments. 
 
\subsection{Noise fluctuations around the optimal local field}

\begin{figure}[t]
	\centering
\includegraphics[width= 1
\columnwidth, ]{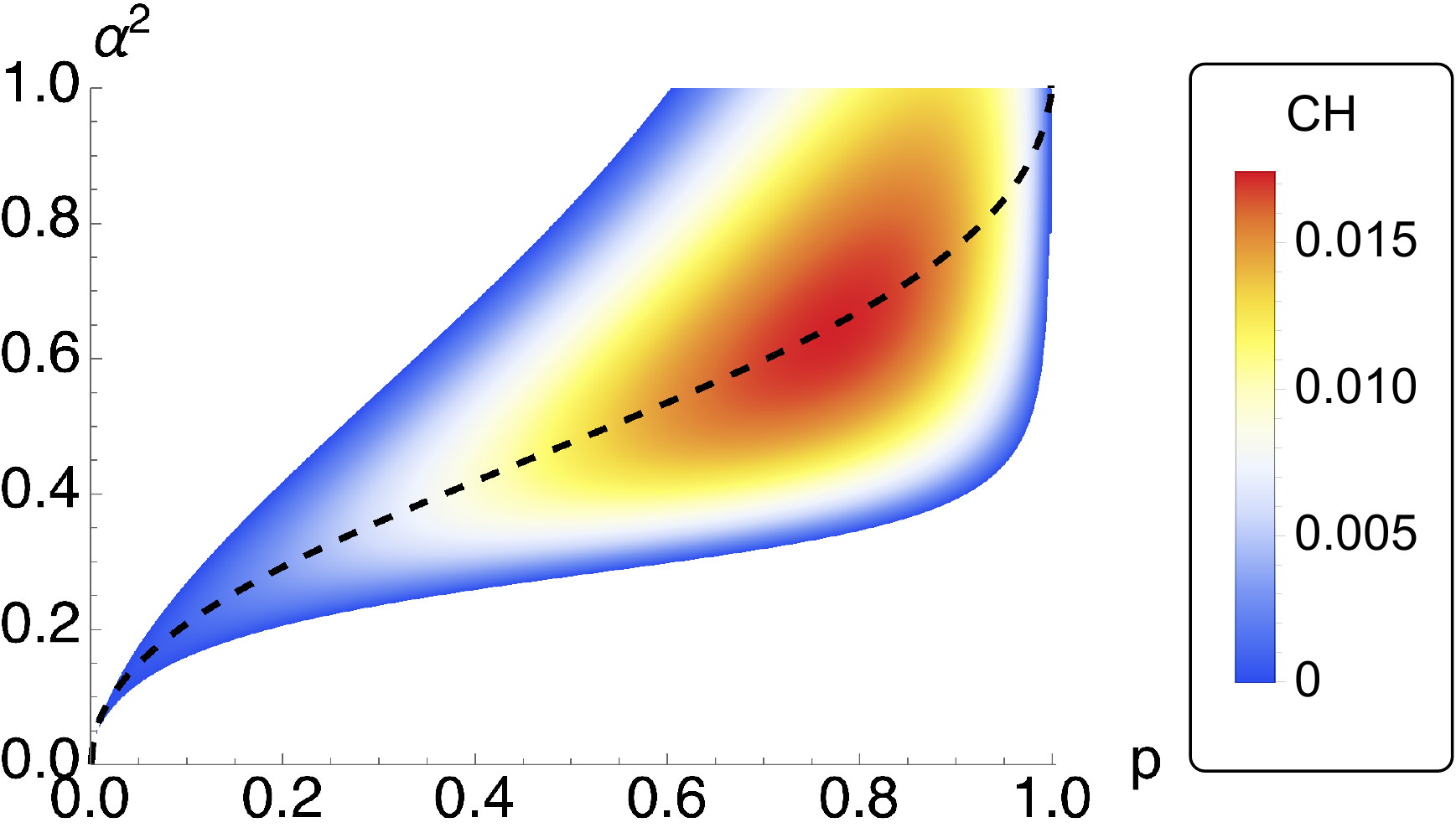}
	\caption{\label{alpharange}
	 Region of violation of the upper bound of the CH inequality (\ref{CHin}) in the plane of $p$ and $\alpha^2$,  for the single-parameter family of quantum states $\ket{\psi(p)}$, as given in (\ref{eq:InState}). $\alpha^2$ represents the intensity of the local oscillator, when the beamsplitter $BS_j$ is \textit{on} for the unprimed settings and it is the same for both the parties. The  reflectivities of the beamsplitters, are taken to be fixed $R_j = \alpha^2_0$, for $j = 1,2$, for \textit{on}, whereas the primed settings are taken as \textit{off}. 
		}
\end{figure}

\begin{figure}[b]
	\centering
\includegraphics[width= 1
\columnwidth, height = 0.62 \columnwidth]{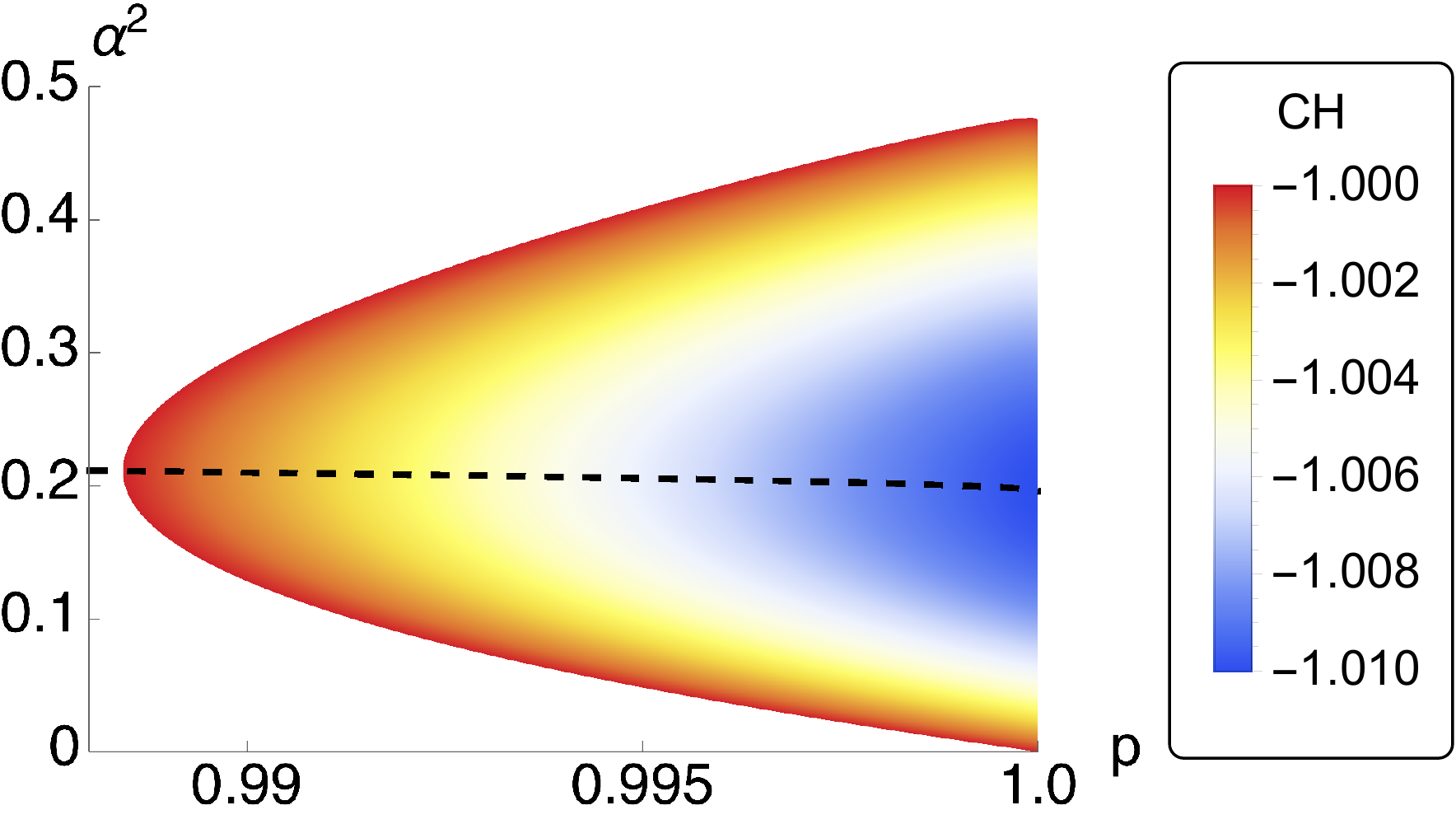}
	\caption{\label{alpharange_min}
		Regionplot of the violation of the CH inequality (\ref{CHin}), in the left hand side, for the state $\ket{\psi(p)}$, for $0.989 < p < 1$ and $\alpha^2 < 1$. Here $\alpha^2$ is the intensity of both the local oscillators, when the beam splitters $BS_j$, of reflectivity $R_j = \alpha^2_0$, for $j = 1,2$, are \textit{on} for the primed settings. The unprimed settings are taken to be \textit{off}. This figure also depicts the sustainable fluctuation  of $\alpha^2$, from the optimum one (the black dotted line), for a fixed $p$, to have violation in the negative side. 
		}
\end{figure}

To address  the effect of  fluctuations of the intensity of the local coherent fields, 
 we checked what is the range of $\alpha^2$ for which it is possible to detect a violation of local realism  while the other measurement settings are fixed to those optimal for the optimal intensity $\alpha_0^2$. The numerical results we obtained this way are presented in Fig. \ref{alpharange} (violation of the upper bound of CH inequality) and Fig. \ref{alpharange_min} (lower bound).  

\begin{figure}[t]
	\centering
\includegraphics[width= 1
\columnwidth, ]{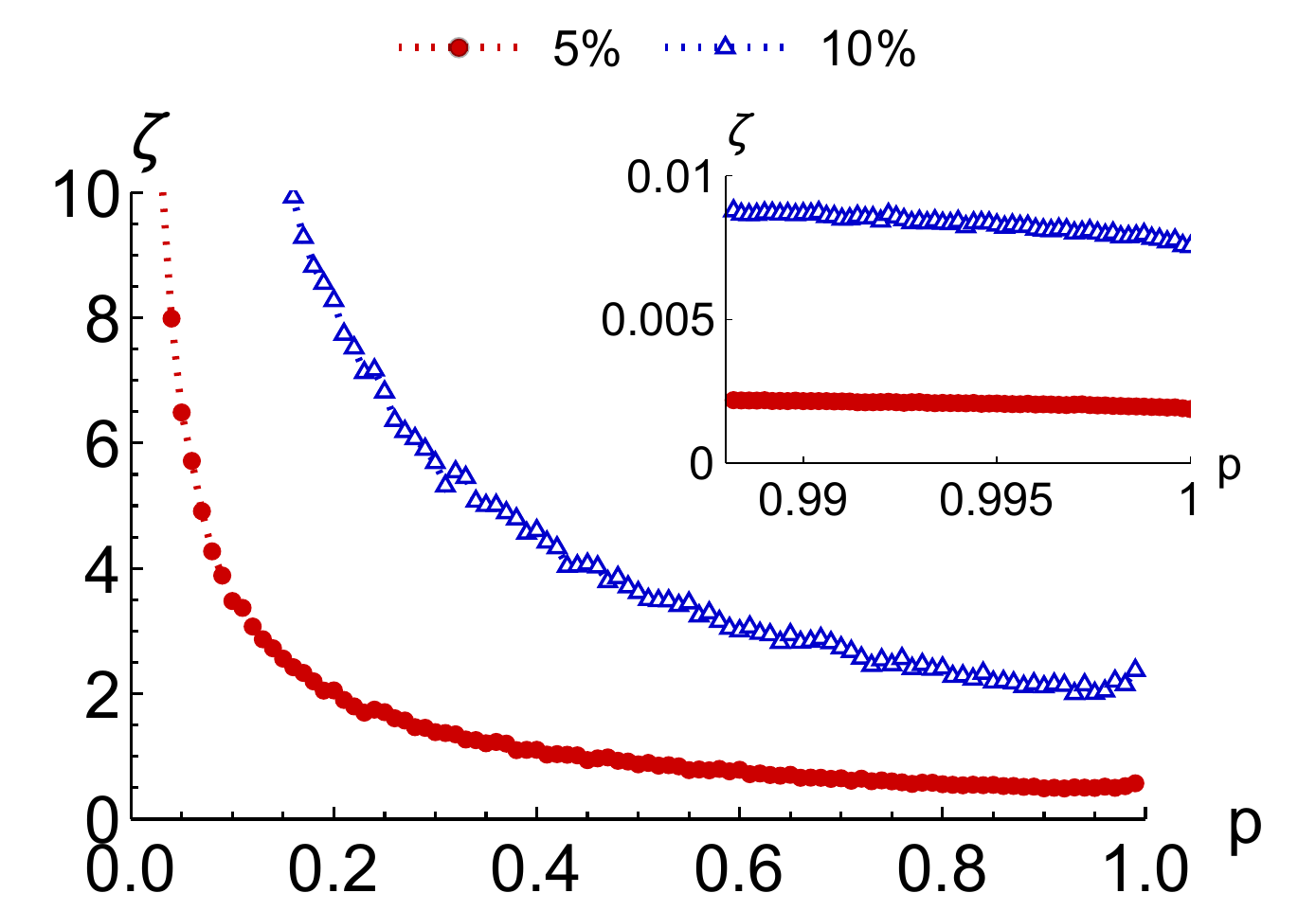}
	\caption{\label{errorCH}
		Plot of $ \%$ of error, $\zeta(p)$, in a simulation of an experimental realization of the violation of CH inequality in the right hand side, with respect to $p$. The red curve is for $\sigma= 0.05 \alpha_0^2$, namely $5\%$ fluctuations of the optimal $ \alpha_0^2$ for each $p$, and the blue one is for $10\%$ fluctuations ($\sigma= 0.1 \alpha_0^2$). 
	The reflectivity of the beam splitter is fixed to the optimal coherent state strength, via $R = \alpha_0^2$. In the inset we have shown the $ \%$ of error in the left hand side violation of (\ref{CHin}), in the close proximity of $p = 1$. As from Fig. \ref{alpharange_min}, it is easy to visualise  that the violation is quite robust around optimal $\alpha_0^2$, and hence the error is very low for left hand side compared to the right hand side violation. The error has been calculated with a sample size of $N = 5000$.}
\end{figure}

We also quantify the impact of the fluctuations of the intensity of local coherent fields $\alpha^2$ on the magnitude of violation of the CH inequality. To this end, we compare the violation obtained for the optimal settings, $CH(\alpha_0^2, p)$, with the one resulting from a detuned intensity of the coherent fields $CH(\alpha^2, p)$ (other measurement settings, like reflectivities, are fixed to those optimal for the $\alpha_0^2$ case).  We define a relative difference of violations as

\begin{equation}
    \zeta(p) \! = \! \int d\alpha^2  \frac{\left|CH(\alpha^2, p)- CH(\alpha_0^2, p)\right|}{CH(\alpha_0^2, p)} \mathcal{N}_{\alpha_0^2, \sigma}(\alpha^2) \times 100\%,
\end{equation}
where $\mathcal{N}_{\alpha_0^2, \sigma}(\alpha^2)$ is a 
truncated normal distribution with a standard deviation $\sigma$  and the lower tail cut off at 0, centered around the optimal intensity $\alpha_0^2$.

The numerical estimates of $\zeta(p)$ for intensity fluctuations proportional to the optimal intensity $\alpha_0^2$ are plotted in Fig. \ref{errorCH}. The violations of the CH inequality prove to be quite robust, especially for $p\approx1$.

\subsection{Inefficient detectors}
Thus far, the photon number resolving detectors we considered were tacitly assumed to have perfect efficiency. In this section, we will investigate detectors of a finite efficiency $\eta < 1$.  
\begin{figure}[t]
	\centering
\includegraphics[width= 1
\columnwidth]{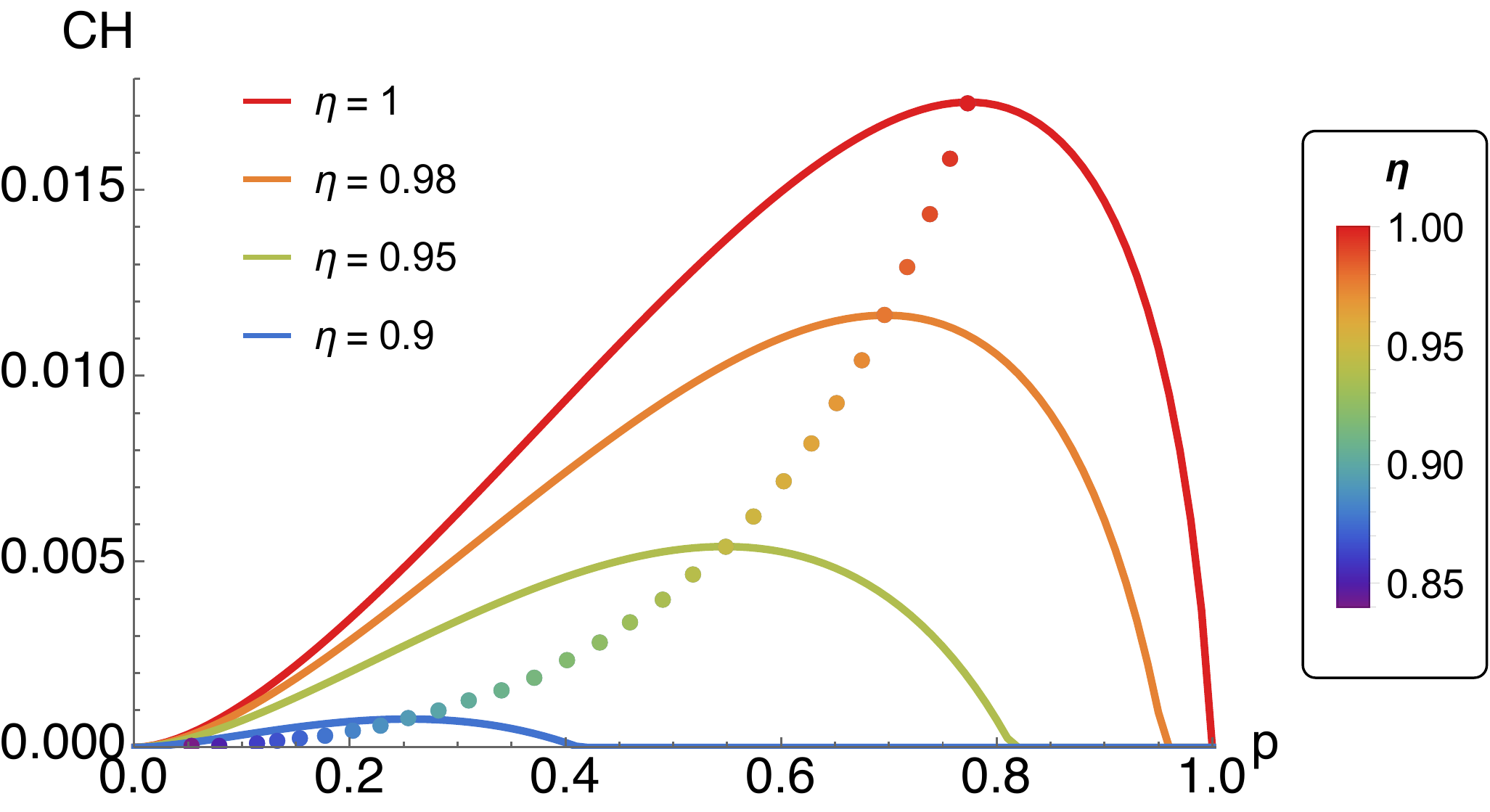}
	\caption{\label{inefficiency}
Maximal violation of CH inequality for different values of detection efficiency $\eta$, as a function of $p$, the probability of having a single photon in the initial state $\ket{\psi(p)}$. The red line corresponds to $\eta = 1$, i.e., for the perfect detectors similar to the plots in figures \ref{CH_max} and \ref{CHallevents}. The orange line is for $\eta = 0.98$
and the  green and blue curves are for $\eta = 0.95$ and $\eta = 0.9$ respectively. The dotted line depicts the migration of the point of maximum CH violation in the plane of $p$ and CH, with various values of $\eta$ (given in the color palette).}
\end{figure}

To do that, we assume that if there are $n$ photons in a given mode,
the detector might not register all of them. Instead, it can detect any $n'$ number of photons, $0 \leq n' \leq n$, with the probability
$\binom{n}{n'} \eta^{n'} (1 - \eta)^{(n - n')}$. Hence, the joint probability of detecting $(k,l;r,s)$ numbers of photons  in modes $c_1, d_1; c_2$, and $d_2$ by detectors with the same efficiency $\eta$, is 
\begin{eqnarray}\label{jointeta}
&&P_\eta(k,l;r,s)_{c_1, d_1, c_2, d_2} = \nonumber  \\
&& \sum_{n,m = 0}^\infty \sum_{n',m' = 0}^\infty \binom{k+n}{k} \binom{l+m}{l} \binom{r+n'}{r} \binom{s+m'}{s} 
 \eta^{k+l + r+s} \nonumber  \\
&& 
 (1 - \eta)^{n+m + n' + m'} P(k+n, l+m, r + n'; r + m')_{c_1, d_1, c_2, d_2}, \nonumber  \\
\end{eqnarray}
 and the local probabilities are
 \begin{eqnarray}\label{localeta}
&&P_\eta(k,l)_{c_1, d_1} = \nonumber  \\
&& \sum_{n,m = 0}^\infty \binom{k+n}{k} \binom{l+m}{l} \eta^{k+l} (1 - \eta)^{n+m} P(k+n, l+m)_{c_1, d_1}, \nonumber  \\
\end{eqnarray}
and the $P_\eta(r,s)_{c_2, d_2}$ have the similar expressions as \eqref{localeta}, see Appendix \ref{inefficient_detector} for more details.
A detailed expressions of the joint and local probabilities of detecting arbitrary number of photons, i.e., detecting $n_j$ number of photons  in mode $c_j$ and $m_j$ in mode $d_j$ for $j = 1,2$ are given in Eqs. (\ref{Pab_n1m1n2m2}) and (\ref{Pa_nm}) of Appendix \ref{appen:arb_prob}.

We calculated the violation of CH inequality, given in \eqref{CHin}, for the set of photon-detection events, in which single photon was detected in mode $d_j$ and no photon was detected in mode $c_j$, i.e., for $k = 0,\, \,l = 1;\,\, r = 0$ and $s = 1$ in Eq. (\ref{jointeta}) and (\ref{localeta}).  
The maximal violation for the one parameter family of vacuum-one-photon qubit $\ket{\psi(p)}$ is plotted in figure \ref{inefficiency} for each $p$ and various values of detector efficiency $\eta$.

Regarding the threshold detector efficiency of our scheme, we assume that it is not possible to experimentally demonstrate a violation of the upper bound of CH inequality of magnitude smaller than $  10^{-5}$. The lowest detection efficiency, for which this violation can be attained, is $\eta \approx 0.844$ (for $p \approx 0.038$).     

 Most importantly, for the optimal violation of CH inequality the measurement settings satisfy the condition $\eta \alpha_j^2 = R_j$ and $\eta {\alpha'_j}^2 = R'_j$,  for both $j = 1,2$.
 Thus the optimality relation holds as before: as the intensity of the local detectors \textit{observed} by the detectors is reduced by  factor of $\eta$. This is an important fact to be taken into account in experimental realizations. It is very interesting that this type of losses does not affect the optimality condition.

\section{Generalizations of the input state}
\subsection{At most single photon states}
\label{sec:generalized}
In the previous sections we have considered the family of input states $\ket{\psi(p)}$, produced by impinging the superposition of vacuum and a single photon on a balanced beamsplitter $BS_0$.
Now suppose that $BS_0$  is instead an unbalanced  beamsplitter of 
transmitivitty $\cos^2 \xi$. This leads to a different family of input states $\ket{\psi(p, \xi)}$ given by
\begin{eqnarray}
\label{eq:InStategen}
   &\ket{\psi(p,\xi)} =&\nonumber\\
   &\left(\sqrt{1-p}\ket{00} + \sqrt{p} \left(\cos \xi \ket{01} +  \sin \xi \ket{10} \right)\right)_{b_1b_2}.&
\end{eqnarray}
These states correspond to a general (up to local phases) situation in which at most a single photon is present in the modes $b_1$ and $b_2$.

To probe the nonclassicality of the states $\ket{\psi(p, \xi)}$ we used the CH inequality (\ref{CHin}) based on single photon detection events described in \ref{detection}. As in the previous sections, we optimized numerically the CH expression over all the measurement settings. 

\begin{figure}[t]
	\centering
\includegraphics[width= 1
\columnwidth]{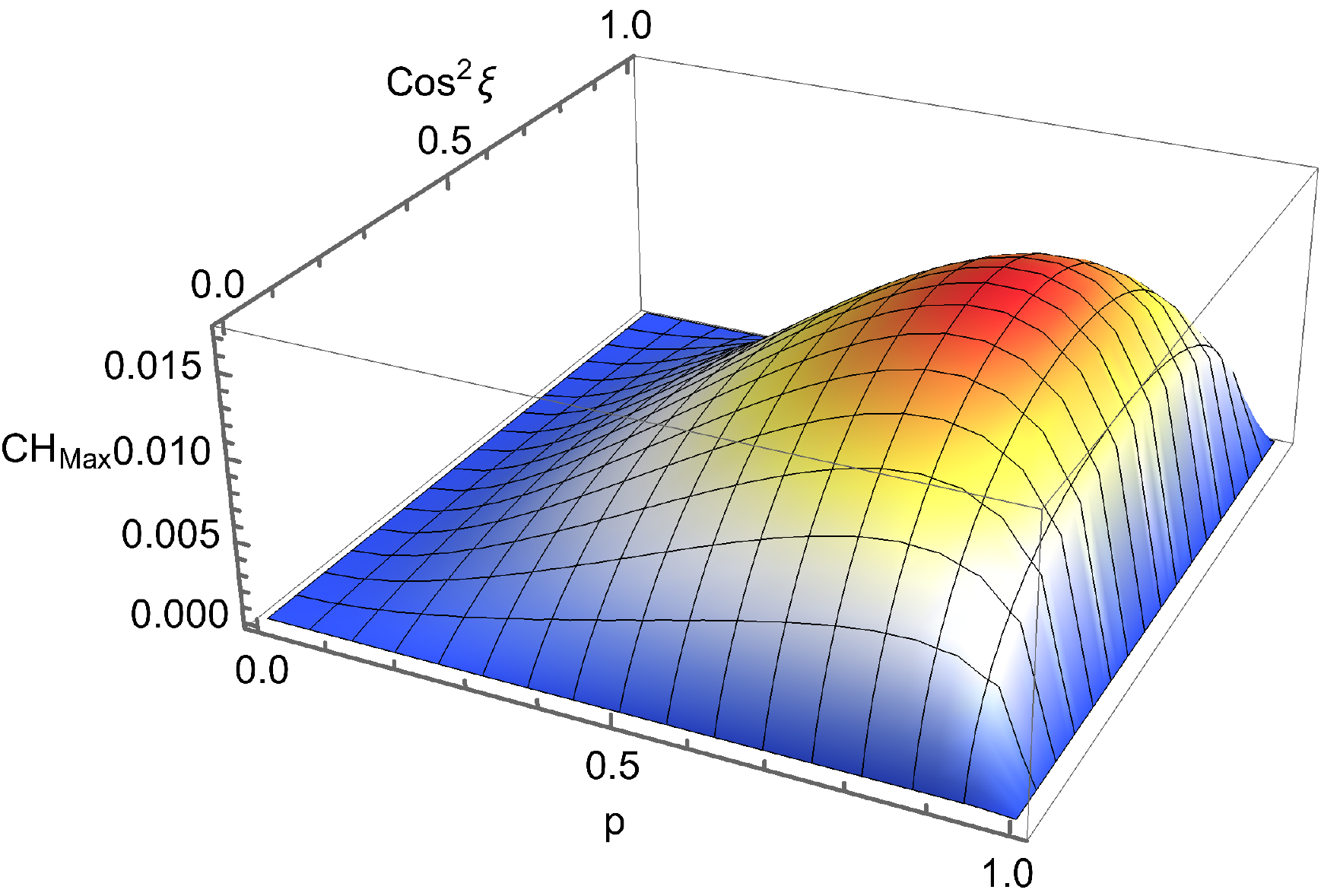}
	\caption{\label{CHmaxgen}
		 Plot of the maximal value $CH_{max}$ of the CH expression for the family of states $\ket{\psi(p,\xi)}$. The optimization was performed  numerically for completely general measurement settings and single-photon detection events (see \ref{detection}). For $0<p<1$ and $0<\xi<\frac{\pi}{2}$ the CH inequality is violated as $CH_{max}>0$.} 
\end{figure}

A violation of the upper bound of (\ref{CHin}), depicted in Fig. \ref{CHmaxgen},  was obtained for the all $0<p<1$ and $0<\xi<\frac{\pi}{2}$. For a given $p$, maximal violation is obtained for $\xi=\frac{\pi}{4}$. The optimal measurement settings which lead to $CH_{max}$ are \textit{on} for unprimed settings (beamsplitters $BS_j$, and the coherent states $\ket{\alpha_j e^{i \phi_j}}$  for $j = 1,2$ are present) and  \textit{off}  for the primed settings (beamsplitters are removed, and local oscillators are turned off). Just as in the case of the balanced $BS_0$, the condition $T_j + \alpha_j^2 = 1$ is satisfied by the optimal \textit{on} measurements for both $j = 1,2$. 
However, the optimal $\alpha_1$ and $\alpha_2$ are not, in general, the same 
when in the initial slate the amplitudes of $\ket{01}_{b_1b_2}$ and $\ket{10}_{b_1b_2}$ are different.

\begin{figure}[t]
	\centering
\includegraphics[width= 1
\columnwidth]{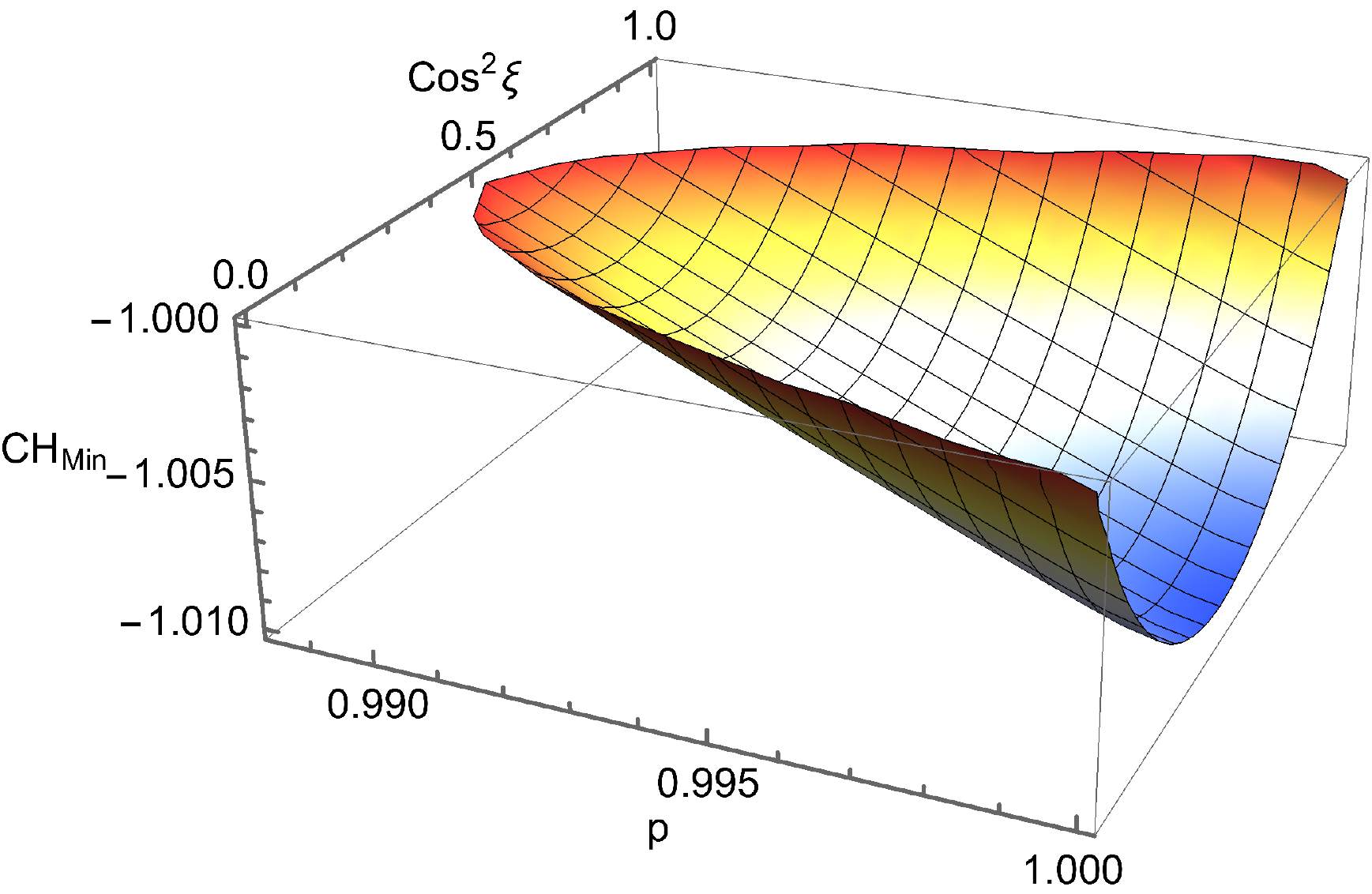}
	\caption{\label{CHmingen}
	Plot of the minimal value $CH_{min}$ of the CH inequality for the family of states $\ket{\psi(p,\xi)}$. The optimization was performed  numerically for completely general measurement settings and single-photon detection events (see \ref{detection}). The violation  has been found for a very small range of parameter $p$. For $p=1$ there is no violation in of the upper bound of the CH inequality, but in the lower bound is violated for the entire range of $\xi$.} 
\end{figure}

Let us look for violations of the lower bound of the CH inequality.  For $p = 1$ the inequality is violated for all $0<\xi<\frac{\pi}{2}$. As $p$ gets smaller, the range of $\xi$ for which a violation can be obtained quickly narrows and finally reduces, from the numerical point of view, to a single point $\xi=\frac{\pi}{4}$ for $p\approx0.989$, see Fig. \ref{CHmingen}. 

The optimal settings are, again, of the \textit{on}/\textit{off} kind and satisfy the condition $T_j + \alpha_j^2 = 1$ for both Alice and Bob.\\


\subsection{$\alpha^2=R$ condition beyond single photon case, a simple example}


\label{sec:TMSV}
The necessary condition for optical measurement settings to violate the CH inequality is puzzling, and thus it begs for an investigation whether it can appear also in other interferometric contexts involving weak homodyne measurements, and single photon detections.
Surprisingly it holds in simple case which we present here. More general studies of this condition in the case off  wider classes of states will be presented somewhere else.

The discussed interferometric configuration  (Fig. \ref{mainSetup}) can be {extended towards investigating other} two-mode input optical states, like  e.g. 
the following one:
\begin{equation}
\label{TMSV}
 \sqrt{1-p} \ket{0, 0}_{b_1,b_2}+ \sqrt p \ket{1, 1}_{b_1,b_2}.
\end{equation}
The joint probability of detecting no photon in mode $c_j$ and single photon in mode $d_j$, for both $j = 1,2$, when both the beamsplitters are present and local oscillators are turned on is:
\begin{eqnarray}
&&P(A,B)= e^{-\alpha_1^2 -\alpha_2^2}\Big[ (1 -p) R_1 R_2 \alpha_1^2 \alpha_2^2 + p (1 - R_1)(1 - R_2) \nonumber \\ 
&&  - 2\alpha_1 \alpha_2 \sqrt{p (1 - p) } \sqrt{R_1(1 - R_1)} \sqrt{R_2(1 - R_2)}\cos(\phi_1 - \phi_2) 
\Big]. \nonumber \\ 
\end{eqnarray}
The local probabilities are:
\begin{equation}
P(A) = e^{-\alpha_1^2} \Big[ (1 - p) R_1 \alpha_1^2 + p (1 - R_1) \Big],
\end{equation}
and 
\begin{equation}
P(B) = e^{-\alpha_2^2} \Big[ (1 - p) R_2 \alpha_2^2 + p (1 - R_2) \Big].
\end{equation}

\begin{figure}[t]
	\centering
\includegraphics[width= 1
\columnwidth]{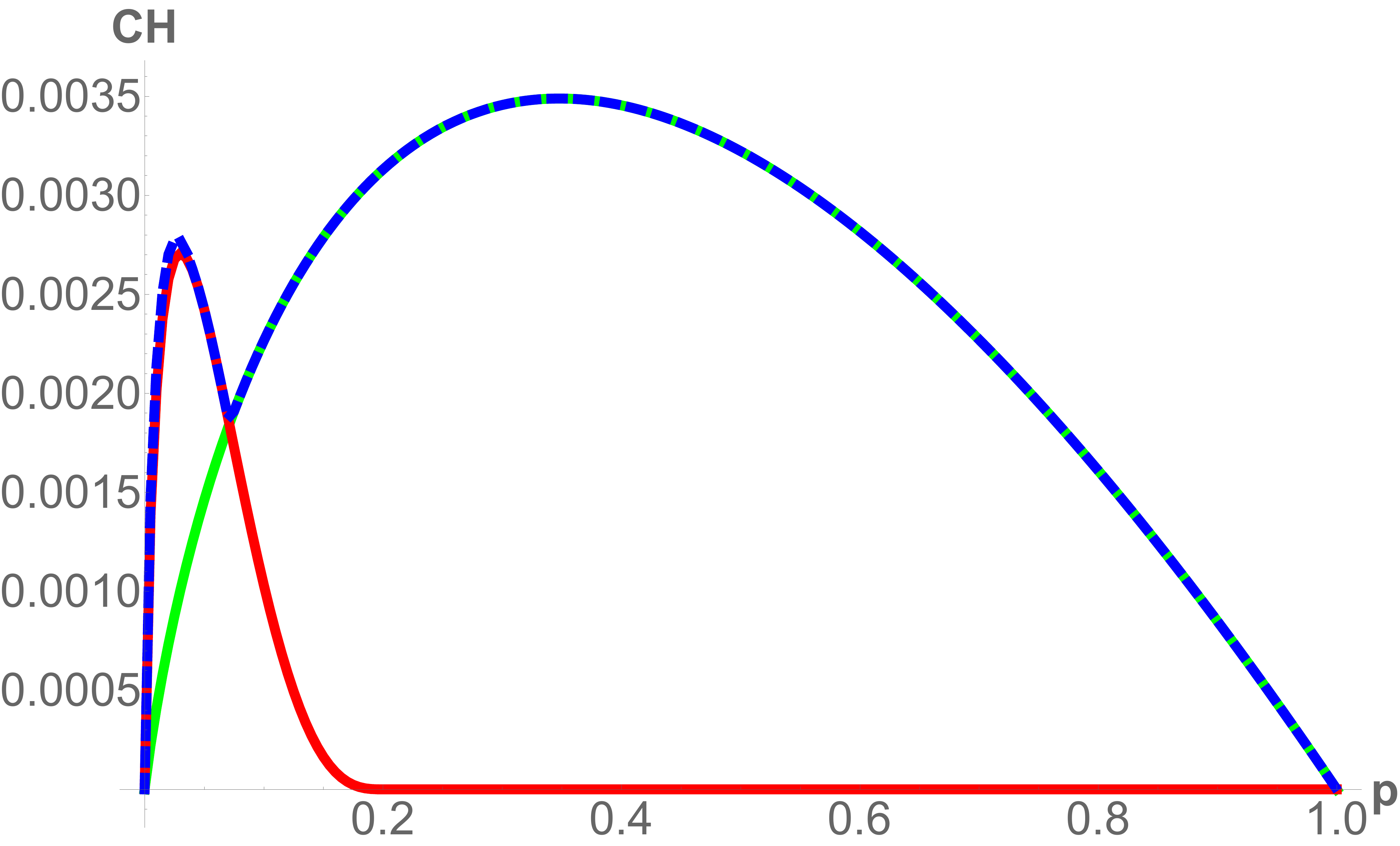}
	\caption{\label{GPY} 
		Plot of the maximal CH violation for the 
		superposition of vacuum and a  two-photon 
		state \eqref{TMSV} as a function of the probability $p$ of having a two-photon component. The red line represents the maximal violation within the perfect \textit{on}/\textit{off} scheme for the  \textit{off} settings chosen as unprimed ones for both observers. The green line represents maximal violation within the perfect \textit{on}/\textit{off} scheme for the  \textit{off} settings chosen as unprimed for one observer and primed  for another. Finally the dashed blue line represents the maximal violation for the general settings (no \textit{on}/\textit{off} constraint). It can be seen that the \textit{on}/\textit{off} settings are almost optimal for lower values of $p$ and exactly optimal for the higher values.}
\end{figure}

Here we  present a numerical optimization of the violations of the CH inequality. 
The  local settings are specified as before by three parameters: amplitude $\alpha$ of the local oscillator, its phase $\phi$ and reflectivity $R$ of the local beamsplitter. 
The events in the CH inequality are defined with respect to single photon detection, as detailed in \ref{detection}. 
It turns out that 
the CH inequality is 
violated for the entire range of $p\in(0,1)$ (see Fig. \ref{GPY}).

For lower values of $p$ the  \textit{on}/\textit{off} settings lead to almost optimal violation (see the red solid line in the Fig. \ref{GPY}), 
if the  \textit{off}  settings correspond to the unprimed measurement choices for both observers. 
On the other hand for higher values of $p$ the exact optimal violation (see green solid line in the Fig. \ref{GPY}) can be obtained if  the  \textit{off}  settings are the primed measurement choices for one observer and unprimed for another (due to the symmetry of the CH expression with respect to a swap of observers it does not matter which one is which).

Surprisingly, both the \textit{on}/\textit{off} optimal settings, and the general optimal ones, outside the \textit{on}/\textit{off} scenario follow the $\alpha_j^2=R_j$ and $\alpha_j'^2=R'_j$ conditions for both primed and unprimed measurement choices of Alice ($j=1$) and Bob ($j=2$). This suggests that the condition $\alpha^2=R$ might be a general property of settings leading to maximal violation of CH inequality based on weak-field homodyne measurements and single-photon detections.

\section{ Closing remarks}

Bell tests are the cornerstone of many quantum protocols, 
device independent quantum key distribution and randomness certification \cite{QRNG_appl, Farkas_2021}. 
Optical states of one or few photons are a feasible choice to implement long-distance protocols in disparate experimental situations, ranging from satellite-based communication \cite{Bell-space} to submarine cable connections \cite{Bell-Submarine}. 
For this reason a detailed study of Bell scenarios  becomes of paramount importance for possible quantum information processing tasks  implemented with such states. All doubtful elements in their analysis must be removed, and the gedanken versions must be translated into ones, which are possible to execute in the laboratories.

Our aim in this work is to move from the foundational level the results obtained in \cite {1stPaper, 2ndPaper}, which show basic inconsistencies in the thus-far offered interpretations of the gedanken-experiments presented in the classic papers \cite{TWC91,GPY}, and turn the discussion into one concerning the conditions required to reveal violations of local realism in laboratory realisations of the experiments.
Our discussion here, and in \cite {1stPaper, 2ndPaper}, is also a basis of re-interpretation of results obtained in current state-of-the-art weak-homodyne photon number resolving experiments, like \cite{Walmsley14, Walmsley20}, which were based on the proposals of \cite{TWC91,GPY}.
Note that in \cite {1stPaper, 2ndPaper}  we have shown, or strongly conjectured, that keeping local oscillator strenghts constant for both local settings, and fixing the transmissivity of the beamsplitters in the detection stations at $T=50\%$, cannot lead to a proper Bell test based on schemes of \cite{TWC91,GPY}, even with detectors of $100\%$ efficiency.

We searched for optimal interferometric scheme revealing {\it true} violation of local realism for a family of initial states $\ket{\psi(p)}$ \eqref{eq:InState}. A scheme proposed in \cite{Hardy94} is a correct Bell test, but is not optimal, as we show here. This is especially so when the vacuum component of the initial state is of a probabilistic weight below $8\%$. Moreover for $p=1$ (ideally a single photon) the scheme does not work. However, note that Hardy did {\it not} aim at the optimality. Still,  our numerical searches and analytic calculations for more tractable cases show that the idea of Hardy for turning {\it off} the local oscillator, and removing the local beamsplitter,  in one of the two local settings on each side is a feature which leads to the optimal violation the CH inequality in the case of local events specified by detection of just one photon in the local measurement station. Also we show that such single-photon detection events (for all settings) are indeed optimal, and that only for $p\approx 1/2$ optimal are  50-50 beamsplitters at the final measurement stations in the {\it on} operation mode. Hardy assumed fixed 50-50 beamsplitters, as it was the case in \cite{TWC91}. 

The most surprising result that we show is that the conditions:
\begin{eqnarray}\label{COND}
    \alpha_j^2 + T_j &=&1  \quad \text{for} \, \, j=1, 2,\nonumber\\
     \alpha_j^2 + R_j &=&1  \quad \text{for} \, \, j=1, 2,
\end{eqnarray}
are necessary to find optimal violation of the CH inequality, when we base our Bell test on single-photon detection events: one photon in mode $d_j$ and no photon in $c_j$ (first condition), or vice versa (second condition). These necessary conditions, despite their beauty, are not easy to interpret. Most importantly they  hold also for the case of inefficient detection in a modified form: 
\begin{equation}
    \eta\alpha_j^2 + T_j =1  \quad \text{for} \, \, j=1, 2,
\end{equation}
and analogously for the second case.
Note that this allows one to  easily choose the optimal settings for the actual experimental setup. Also, we find an interesting situation, which is one of the characteristic trait of the experiment: optimal settings change with the efficiency of the detection. 

Surprisingly, the conditions (\ref{COND}) are also necessary ones for optimality of settings in the case of state (\ref{TMSV}). In a forthcoming article we discuss a modified experiment of \cite{GPY}, which involves weak homodyne measurement on two mode (beam) squeezed vacuum. The modification rest on \textit{on}/\textit{off} settings and tunable beamsplitters at the measurement stations, for single-photon detection events for each party. Thus, the conditions might be some general feature of weak homodyne measurements in Bell tests. This will be discussed in another article. An open question is to find the entire family of states for which the optimality conditions (\ref{COND}) hold.

\section*{Acknowledgements}
This work is supported by  Foundation for Polish Science (FNP), IRAP project ICTQT, contract no. 2018/MAB/5, co-financed by EU  Smart Growth Operational Programme. MK is supported by  FNP  START scholarship. AM is supported by (Polish) National Science Center (NCN): MINIATURA  DEC-2020/04/X/ST2/01794.

 \appendix 
 
\section{Hardy's argument vs ours}
\label{s:hardyexp}
 When discussing here the version of the gedanken-experiment by Hardy, we shall use a slightly different  initial state of beams $b_j$, namely
 \begin{equation}
\label{eq:InStateHARDY}
  \sqrt{1-p}\ket{00}_{b_1,b_2} + \sqrt{\frac{p}{2}} \left(\ket{01}_{b_1,b_2} + i\ket{10}_{b_1,b_2} \right).
\end{equation}
This state was used by Hardy.
 It differs form, $\ket{\psi(p)}$, formula (\ref{eq:InState}), by a trivial $\pi/2$  phase shift in beam $b_1$.
 We decided to use in the main text (\ref{eq:InState}) because in its a case the formulas for probabilities become symmetric with respect an Alice-Bob interchange, and thus also the optimal settings acquire a fully symmetric form.
 
Hardy \cite{Hardy94}, considered the following measurement settings for both Alice and Bob.

\textit{ Local settings $U_j$:}  No beamsplitter in mode $b_j$ or the beamsplitter BS$_j$ has $100\%$  transmittance. If single photon is detected in detector ${d_j}$, (in this case we have $b_j \equiv d_j$) then the corresponding outcome is considered as $U_j = 1$ otherwise it is $U_j = 0$. The definition of the $U$ event is effectively the same, if one additionally switches  \textit{off} the local oscillator, as local oscillator photons cannot reach the $d$ detectors when the beamsplitter is removed. Thus this definition is concurrent with our for the \textit{off} setting.

\textit{Local settings $F_j$:} The beamsplitter BS$_j$,  is a 50-50 one, and the input state $\ket{\psi(p)}$, interferes with  auxiliary coherent  beams $\ket{\alpha_j}$, for $j = 1,2$.  
If precisely a single photon is detected in ${d_j}$ and no photon clicks in ${c_j}$, then the event is put by Hardy as $F_j = 1$.  This is also concurrent with our definition of the \textit{on} setting and the result considered by us is $F_j=1$.

\textit{CH inequality:} 
The reasoning given by Hardy can be linked with CH inequalities in the following way. 
The following CH inequality must hold for the considered events:
\begin{eqnarray}\label{eq:Hardy_CH0}
  &&  -1 \leq P(F_1 = 1, F_2 = 1) + P(F_1 = 1, U_2 = 1) \nonumber \\ 
  && \hspace{0.65cm} + P(U_1 = 1, F_2 = 1) 
    - P(U_1 = 1, U_2 = 1) \nonumber \\
  && \hspace{2cm} - P(F_1 = 1) - P(F_2 = 1)\leq 0,
\end{eqnarray}
where $P(X_1 = 1, Y_2 = 1)$,  denotes the probability of getting outcomes $X_1 = 1$, and $Y_2 = 1$ of joint measurements $X_1$ and $Y_2$, by two spacelike separated observers Alice and Bob, for $X,Y \in \{U, F\}$.
The above inequality is equivalent to
\begin{eqnarray}\label{eq:Hardy_CH}
    && -1 \leq P(F_1 = 1, F_2 = 1) - P(F_1 = 1, U_2 = 0) \nonumber \\
    && \hspace{0.65cm} - P(U_1 = 0, F_2 = 1) - P(U_1 = 1, U_2 = 1) \leq 0.\nonumber\\
 \end{eqnarray}
 This can be shown using the fact that $P(A) - P(A, B)= P(A, \bar{B})$,  where $\bar{B}$  is the opposite event with respect to $B$. This trick is done here  for events $U_j$. The opposite event to $U_j=1$ is $U_j=0$, which in fact means no photon detected in $d_j$.

The right hand side of the new inequality can be put as follows
\begin{eqnarray}\label{eq:Hardy_CH-2}
     P(F_1 = 1, F_2 = 1) \leq P(F_1 = 1, U_2 = 0) \nonumber \\
     + P(U_1 = 0, F_2 = 1) + P(U_1 = 1, U_2 = 1). 
 \end{eqnarray}
This is equivalent to the Hardy's paradox: in the local realistic case if the three right-hand-side probabilities are zero, then the  left hand side one must be zero.

In the quantum case one seeks situations in which  all three right hand side probabilities are zero and we have $P_{quantum}(F_1 = 1, F_2 = 1)>0.$
Hardy obtained the non-zero $P_{quantum}$ for his settings:
\begin{equation}
   CH_{Hardy} = P_{quantum}(F_1 = 1, F_2 = 1) =  \frac{e^{-\frac{p}{1-p}} p^2}{16 (1-p)}.
\end{equation}
He chooses the input coherent state amplitudes, their phases are now included, of the values   $\alpha_1 = - \sqrt{\frac{p}{2(1-p)}} $ for mode ${a_1}$
and $\alpha_2 = i\sqrt{\frac{p}{2(1-p)}} $ for ${a_2}$.  This is so for his $F$ settings, our \textit{on} ones. This makes the other three probabilities vanish.
Probability $P_{quantum}(F_1 = 1, F_2 = 1)$ gives the value of the CH expression (\ref{CHin}), for the Hardy approach.
A plot of it is given in Fig. \ref{CH_max}.

Fig. \ref{CH_max} shows that, in  $p$  close to $1$  the value of $CH_{Hardy}$ is minuscule. We have, $P_{quantum} < 10^{-6}$, for $p > 0.922$. Thus with the approach it is experimentally impossible to detect the non-classicality of the single photon state for such values of $p$.  Still, the ideal prediction  $P_{quantum} > 0$ holds for  the entire range of $0<p<1$, and hence, in principle Hardy's approach can detect the non-classicality of the state (\ref{eq:InState}) for any $p$ within the range, however, tellingly, not for $p=1$. 

\begin{figure}[b]
	\centering
\includegraphics[width= 0.9
\columnwidth]{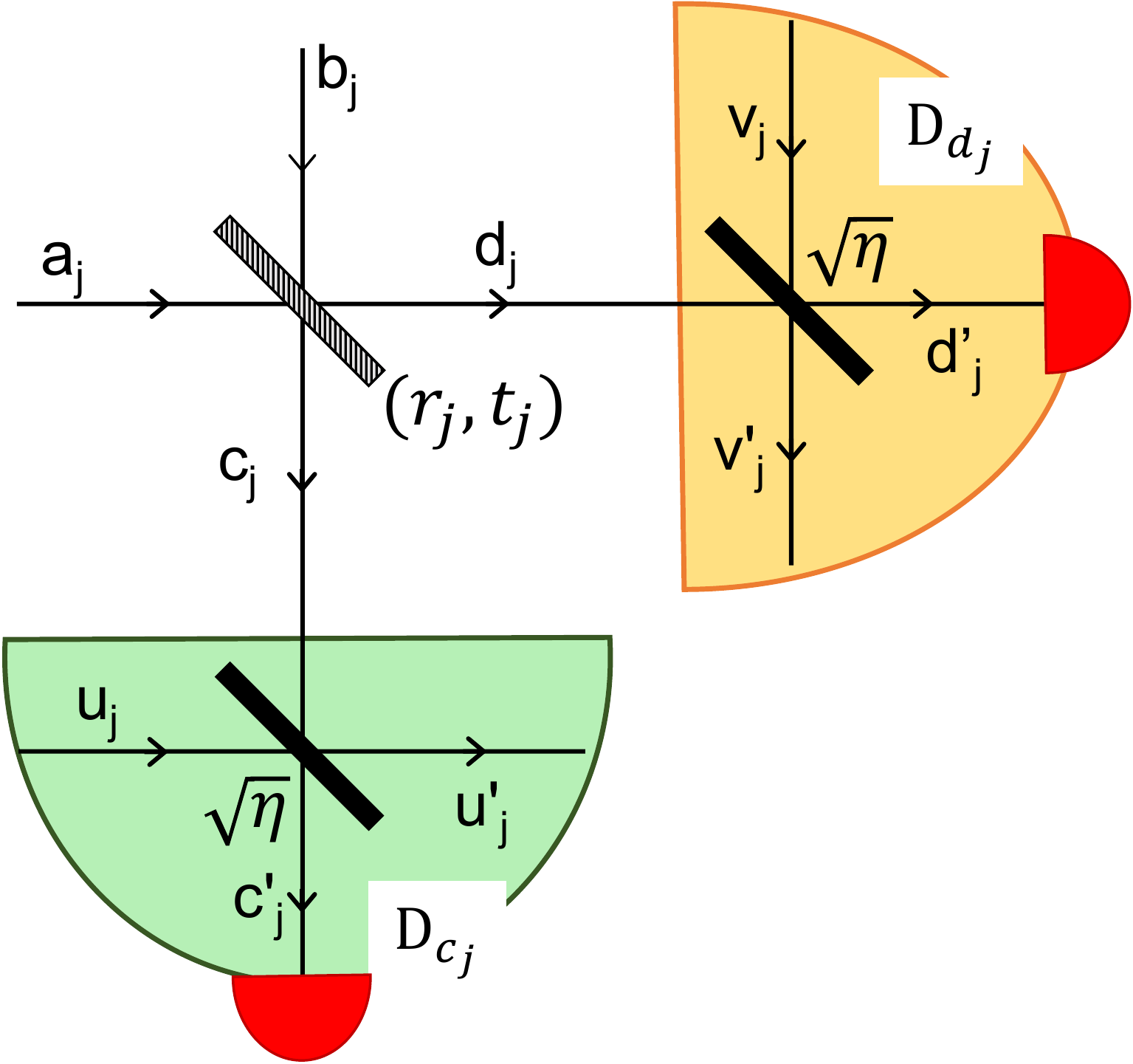}
	\caption{\label{ineffi_detector}
		Schematic diagram of inefficient detector.
		}
\end{figure}
All that was said above points to the fact that the method directly employing the Hardy paradox used a constrained version of the CH inequality, and thus is less effective in detection of local realism than the one which we present in the main text.
This is not a criticism of Hardy's result, as one of his aims was to show an application of his paradox.

\section{Modeling detector inefficiency}\label{inefficient_detector}
We modeled the  inefficiency of a   detector,  by 
introducing \textit{imaginary} additional beam splitter of transmissivity $\eta$ in  the modes $c_j$ and $d_j$. 
 The two other input modes $u_j$ and $v_j$ of the  imaginary beam splitters, as shown in the figure \ref{ineffi_detector}, then one has
\begin{eqnarray}
\begin{pmatrix}
\hat x_j \\ \hat y_j,
\end{pmatrix} = 
\begin{pmatrix}
\sqrt{\eta} & \sqrt{1 - \eta} \\
-\sqrt{1 - \eta} & \sqrt{\eta}
\end{pmatrix}. 
\begin{pmatrix}
\hat x'_j \\ \hat y'_j,
\end{pmatrix}
\end{eqnarray}
where $x = c,d$ and $y = u, v$, and Similarly $u'_j$ and $v'_j$ are  the loss modes of the imaginary beam splitters modelling the inefficiency.


Now, the joint probability of getting $(k,l;r,s)$ photons in mode $c'_1, d'_1; c'_2, d'_2$ is given by
\begin{widetext}
\begin{eqnarray}
P(k,l;r,s)_{c'_1, d'_1; c'_2, d'_2} =  \text{tr}\big(\ket{k,l}\bra{k,l}_{c'_1, d'_1} \otimes \ket{r,s}\bra{r,s}_{c'_2, d'_2} \otimes I_{u'_1v'_1} \otimes I_{u'_2v'_2} \ket{\Psi_{in}(p)}\bra{\Psi_{in}(p)}\big),
\end{eqnarray}
and the local probability is 
\begin{eqnarray}\label{eq:local_loss_prob}
P(k,l)_{c'_1, d'_1} =  \text{tr}\big(\ket{k,l}\bra{k,l}_{c'_1, d'_1}  \otimes I_{u'_1v'_1}  \ket{\Psi_{in}(p)}\bra{\Psi_{in}(p)}\big),
\end{eqnarray}
\end{widetext}
where $I_{u'_jv'_j}$, for $j = 1,2$ are the identity operators in the  loss modes.  One can write Eq. (\ref{eq:local_loss_prob}) as
\begin{eqnarray}
P(k,l)_{c'_1, d'_1} =  \sum_{n,m = 0}^\infty\big|(\bra{k,l}_{c'_1, d'_1}  \otimes \bra{n,m}_{u'_1v'_1} ) \ket{\Psi_{in}(p)}\big|^2,~~~~~~
\end{eqnarray}
where we use the fact that $I_{u'_1v'_1}  = \sum_{n,m = 0}^\infty \ket{n,m}\bra{n,m}_{u'_1v'_1}$. 

Now expand the state $\ket{k,l}_{c'_1, d'_1}  \otimes \ket{n,m}_{u'_1v'_1}$ in terms of the input modes 
\begin{eqnarray}
&& \ket{k,l}_{c'_1, d'_1}  \otimes \ket{n,m}_{u'_1v'_1} \nonumber \\
&& = \frac{1}{\sqrt{k! l!  n! m!}} ({c'}^\dagger_1)^k ({d'}^\dagger_1)^l ({u'}^\dagger_1)^n ({v'}^\dagger_1)^m \ket{\Omega}~~~~~~ \\
&& = \frac{1}{\sqrt{k! l!  n! m!}} (\sqrt{\eta}~ c_1^\dagger + \sqrt{1 - \eta} ~u_1^\dagger)^k (\sqrt{\eta} ~d_1^\dagger + \sqrt{1 - \eta}~ v_1^\dagger)^l \nonumber \\
&& (\sqrt{\eta} ~u_1^\dagger - \sqrt{1 - \eta}~ c_1^\dagger)^n (\sqrt{\eta} ~v_1^\dagger - \sqrt{1 - \eta} ~d_1^\dagger)^m \ket{\Omega}~~~~~ \label{eq:loss_ex1}\\
&& = \frac{(-1)^{m+n}}{\sqrt{k! l!  n! m!}}~ \eta^{\frac{k+l}{2}} (1 -\eta)^{\frac{n+m}{2}} (c_1^\dagger)^{k+n} (c_1^\dagger)^{l+m} \ket{\Omega} \label{eq:loss_ex2}\\
&& = \frac{(-1)^{m+n}}{\sqrt{k! l!  n! m!}}~ \eta^{\frac{k+l}{2}} (1 -\eta)^{\frac{n+m}{2}} \sqrt{(k+n)!} \sqrt{(l+m)!} \nonumber \\
&& \hspace{1.2in} \ket{k+n, l+m}_{c_1, d_1} \ket{\Omega}_{u_1v_1}.
\end{eqnarray}
From Eq. (\ref{eq:loss_ex1}) to (\ref{eq:loss_ex2}), we use the  feature of the loss model that there is vacuum in modes $u_1$ and $v_1$, as in figure \ref{ineffi_detector}, hence any terms containing  $\hat u_1^\dagger$ and $\hat v_1^\dagger$ simply vanishes when sandwiched with $\ket{\Psi_{in}(p)}$. Hence, 
Eq. (\ref{eq:local_loss_prob}), reduces to 
\begin{eqnarray}
&& P(k,l)_{c'_1, d'_1} =  \sum_{n,m = 0}^\infty \binom{k+n}{k} \binom{l+m}{l} \nonumber \\
&& \hspace{1in}
\eta^{k+l} (1 - \eta)^{n+m} P(k+n, l+m)_{c_1, d_1},~~~~~~
\end{eqnarray}
now put $P(k,l)_{c'_1, d'_1} \equiv P_{\eta}(k,l)_{c_1, d_1}$, we obtained Eq. (\ref{localeta}), and similarly Eq. (\ref{jointeta}).

\section{Probabilities for arbitrary number of photo-detection events}\label{appen:arb_prob}

The joint probability of detecting arbitrary number of photons, where $n_j$ number of photons have been detected in mode $c_j$ and $m_j$ number of photons in mode $d_j$ for $j = 1,2$ is given by

\begin{eqnarray}
&& P(n_1, m_1; n_2,m_2)_{c_1,d_1;c_2, d_2} = \frac{e^{-\alpha _1^2-\alpha _2^2}}{m_1! m_2! n_1! n_2!} R_1^{m_1-1} R_2^{m_2-1} \times \nonumber \\
&&  
\hspace{1em} \left(1-R_1\right){}^{n_1-1} \left(1-R_2\right){}^{n_2-1} \alpha _1^{2 \left(m_1+n_1-1\right)} \alpha _2^{2 \left(m_1+n_1-1\right)} \times \nonumber \\
&& \hspace{0.5em}
\Big( \alpha _1^2 \alpha _2^2 (1-p) \left(1-R_1\right) R_1 \left(1-R_2\right) R_2 \nonumber \\
&& \hspace{1em}
+~ \frac{p}{2} \alpha _1^2  \left(1-R_1\right) R_1 \left(m_2 \left(1-R_2\right)-n_2 R_2\right){}^2  \nonumber \\
&& \hspace{1em}
+~ \frac{p}{2} \alpha _2^2  \left(1-R_2\right) R_2 \left(m_1 \left(1-R_1\right)-n_1 R_1\right){}^2  \nonumber \\
&& \hspace{1em}
- \sqrt{2} \alpha _1 \alpha _2^2 \sqrt{p(1-p)}  \sqrt{R_1(1-R_1)}  \left(1-R_2\right) R_2  \times \nonumber \\
&& \hspace{12em}
\left(m_1 \left(1-R_1\right)-n_1 R_1\right)\sin (\phi_1) \nonumber \\
&& \hspace{1em}
- \sqrt{2} \alpha _1^2 \alpha _2 \sqrt{p(1-p)} \left(1-R_1\right) R_1  
\sqrt{R_2(1-R_2)}  \times \nonumber \\
&& \hspace{12em}
\left(m_2 \left(1-R_2\right)-n_2 R_2\right)\sin (\phi_2) \nonumber \\
&& \hspace{1em}
+~ p \alpha _1 \alpha _2  \sqrt{R_1(1-R_1)} \sqrt{R_2(1-R_2)} \cos (\phi_1 - \phi_2) \times \nonumber \\
&&
\hspace{3em} 
\left(m_1 \left(1-R_1\right)-n_1 R_1\right) \left(m_2 \left(1-R_2\right)-n_2 R_2\right) \Big). \label{Pab_n1m1n2m2}
\end{eqnarray}

The local probability for Alice is
\begin{eqnarray}\label{Pa_nm}
&& P(n,m)_{c_1,d_1} = 
\frac{1}{m! n!}e^{-\alpha _1^2} R_1{}^{m-1} (1-R_1)^{n-1} \alpha _1^{2 (m+n-1)} \nonumber \\
&& \hspace{-1.3em} \Big(-\sqrt{2} \alpha _1 \sqrt{(1-p) p} \sqrt{\left(1-R_1\right) R_1} \sin \left(\phi _1\right) \left(m \left(1-R_1\right) -n R_1\right) \nonumber \\
&& +~ \frac{p}{2}  \left(m \left(1-R_1\right)-n R_1\right){}^2+\frac{1}{2} \alpha _1^2 (2-p) \left(1-R_1\right) R_1\Big). 
\end{eqnarray}
The local probability for Bob, $P(n,m)_{c_2,d_2}$, is exactly same as (\ref{Pa_nm}), with $1 \leftrightarrow 2$.


\begin{thebibliography}{32}%
\makeatletter
\providecommand \@ifxundefined [1]{%
 \@ifx{#1\undefined}
}%
\providecommand \@ifnum [1]{%
 \ifnum #1\expandafter \@firstoftwo
 \else \expandafter \@secondoftwo
 \fi
}%
\providecommand \@ifx [1]{%
 \ifx #1\expandafter \@firstoftwo
 \else \expandafter \@secondoftwo
 \fi
}%
\providecommand \natexlab [1]{#1}%
\providecommand \enquote  [1]{``#1''}%
\providecommand \bibnamefont  [1]{#1}%
\providecommand \bibfnamefont [1]{#1}%
\providecommand \citenamefont [1]{#1}%
\providecommand \href@noop [0]{\@secondoftwo}%
\providecommand \href [0]{\begingroup \@sanitize@url \@href}%
\providecommand \@href[1]{\@@startlink{#1}\@@href}%
\providecommand \@@href[1]{\endgroup#1\@@endlink}%
\providecommand \@sanitize@url [0]{\catcode `\\12\catcode `\$12\catcode
  `\&12\catcode `\#12\catcode `\^12\catcode `\_12\catcode `\%12\relax}%
\providecommand \@@startlink[1]{}%
\providecommand \@@endlink[0]{}%
\providecommand \url  [0]{\begingroup\@sanitize@url \@url }%
\providecommand \@url [1]{\endgroup\@href {#1}{\urlprefix }}%
\providecommand \urlprefix  [0]{URL }%
\providecommand \Eprint [0]{\href }%
\providecommand \doibase [0]{https://doi.org/}%
\providecommand \selectlanguage [0]{\@gobble}%
\providecommand \bibinfo  [0]{\@secondoftwo}%
\providecommand \bibfield  [0]{\@secondoftwo}%
\providecommand \translation [1]{[#1]}%
\providecommand \BibitemOpen [0]{}%
\providecommand \bibitemStop [0]{}%
\providecommand \bibitemNoStop [0]{.\EOS\space}%
\providecommand \EOS [0]{\spacefactor3000\relax}%
\providecommand \BibitemShut  [1]{\csname bibitem#1\endcsname}%
\let\auto@bib@innerbib\@empty
\bibitem [{\citenamefont {Horodecki}\ \emph {et~al.}(2009)\citenamefont
  {Horodecki}, \citenamefont {Horodecki}, \citenamefont {Horodecki},\ and\
  \citenamefont {Horodecki}}]{HHH09}%
  \BibitemOpen
  \bibfield  {author} {\bibinfo {author} {\bibfnamefont {R.}~\bibnamefont
  {Horodecki}}, \bibinfo {author} {\bibfnamefont {P.}~\bibnamefont
  {Horodecki}}, \bibinfo {author} {\bibfnamefont {M.}~\bibnamefont
  {Horodecki}},\ and\ \bibinfo {author} {\bibfnamefont {K.}~\bibnamefont
  {Horodecki}},\ }\bibfield  {title} {\bibinfo {title} {Quantum entanglement},\
  }\href {https://doi.org/10.1103/RevModPhys.81.865} {\bibfield  {journal}
  {\bibinfo  {journal} {Rev. Mod. Phys.}\ }\textbf {\bibinfo {volume} {81}},\
  \bibinfo {pages} {865} (\bibinfo {year} {2009})}\BibitemShut {NoStop}%
\bibitem [{\citenamefont {Pan}\ \emph {et~al.}(2012{\natexlab{a}})\citenamefont
  {Pan}, \citenamefont {Chen}, \citenamefont {Lu}, \citenamefont {Weinfurter},
  \citenamefont {Zeilinger},\ and\ \citenamefont {\ifmmode~\dot{Z}\else
  \.{Z}\fi{}ukowski}}]{ZukowskiRMP}%
  \BibitemOpen
  \bibfield  {author} {\bibinfo {author} {\bibfnamefont {J.-W.}\ \bibnamefont
  {Pan}}, \bibinfo {author} {\bibfnamefont {Z.-B.}\ \bibnamefont {Chen}},
  \bibinfo {author} {\bibfnamefont {C.-Y.}\ \bibnamefont {Lu}}, \bibinfo
  {author} {\bibfnamefont {H.}~\bibnamefont {Weinfurter}}, \bibinfo {author}
  {\bibfnamefont {A.}~\bibnamefont {Zeilinger}},\ and\ \bibinfo {author}
  {\bibfnamefont {M.}~\bibnamefont {\ifmmode~\dot{Z}\else \.{Z}\fi{}ukowski}},\
  }\bibfield  {title} {\bibinfo {title} {Multiphoton entanglement and
  interferometry},\ }\href {https://doi.org/10.1103/RevModPhys.84.777}
  {\bibfield  {journal} {\bibinfo  {journal} {Rev. Mod. Phys.}\ }\textbf
  {\bibinfo {volume} {84}},\ \bibinfo {pages} {777} (\bibinfo {year}
  {2012}{\natexlab{a}})}\BibitemShut {NoStop}%
\bibitem [{\citenamefont {Brunner}\ \emph {et~al.}(2014)\citenamefont
  {Brunner}, \citenamefont {Cavalcanti}, \citenamefont {Pironio}, \citenamefont
  {Scarani},\ and\ \citenamefont {Wehner}}]{Brunner14}%
  \BibitemOpen
  \bibfield  {author} {\bibinfo {author} {\bibfnamefont {N.}~\bibnamefont
  {Brunner}}, \bibinfo {author} {\bibfnamefont {D.}~\bibnamefont {Cavalcanti}},
  \bibinfo {author} {\bibfnamefont {S.}~\bibnamefont {Pironio}}, \bibinfo
  {author} {\bibfnamefont {V.}~\bibnamefont {Scarani}},\ and\ \bibinfo {author}
  {\bibfnamefont {S.}~\bibnamefont {Wehner}},\ }\bibfield  {title} {\bibinfo
  {title} {Bell nonlocality},\ }\href
  {https://doi.org/10.1103/RevModPhys.86.419} {\bibfield  {journal} {\bibinfo
  {journal} {Rev. Mod. Phys.}\ }\textbf {\bibinfo {volume} {86}},\ \bibinfo
  {pages} {419} (\bibinfo {year} {2014})}\BibitemShut {NoStop}%
\bibitem [{\citenamefont {Tan}\ \emph {et~al.}(1991)\citenamefont {Tan},
  \citenamefont {Walls},\ and\ \citenamefont {Collett}}]{TWC91}%
  \BibitemOpen
  \bibfield  {author} {\bibinfo {author} {\bibfnamefont {S.~M.}\ \bibnamefont
  {Tan}}, \bibinfo {author} {\bibfnamefont {D.~F.}\ \bibnamefont {Walls}},\
  and\ \bibinfo {author} {\bibfnamefont {M.~J.}\ \bibnamefont {Collett}},\
  }\bibfield  {title} {\bibinfo {title} {Nonlocality of a single photon},\
  }\href {https://doi.org/10.1103/PhysRevLett.66.252} {\bibfield  {journal}
  {\bibinfo  {journal} {Phys. Rev. Lett.}\ }\textbf {\bibinfo {volume} {66}},\
  \bibinfo {pages} {252} (\bibinfo {year} {1991})}\BibitemShut {NoStop}%
\bibitem [{\citenamefont {Hardy}(1994)}]{Hardy94}%
  \BibitemOpen
  \bibfield  {author} {\bibinfo {author} {\bibfnamefont {L.}~\bibnamefont
  {Hardy}},\ }\bibfield  {title} {\bibinfo {title} {Nonlocality of a single
  photon revisited},\ }\href {https://doi.org/10.1103/PhysRevLett.73.2279}
  {\bibfield  {journal} {\bibinfo  {journal} {Phys. Rev. Lett.}\ }\textbf
  {\bibinfo {volume} {73}},\ \bibinfo {pages} {2279} (\bibinfo {year}
  {1994})}\BibitemShut {NoStop}%
\bibitem [{\citenamefont {Vaidman}(1995)}]{Vaidman95}%
  \BibitemOpen
  \bibfield  {author} {\bibinfo {author} {\bibfnamefont {L.}~\bibnamefont
  {Vaidman}},\ }\bibfield  {title} {\bibinfo {title} {Nonlocality of a single
  photon revisited again},\ }\href
  {https://doi.org/10.1103/PhysRevLett.75.2063} {\bibfield  {journal} {\bibinfo
   {journal} {Phys. Rev. Lett.}\ }\textbf {\bibinfo {volume} {75}},\ \bibinfo
  {pages} {2063} (\bibinfo {year} {1995})}\BibitemShut {NoStop}%
\bibitem [{\citenamefont {Hessmo}\ \emph {et~al.}(2004)\citenamefont {Hessmo},
  \citenamefont {Usachev}, \citenamefont {Heydari},\ and\ \citenamefont
  {Bj\"ork}}]{Hessmo04}%
  \BibitemOpen
  \bibfield  {author} {\bibinfo {author} {\bibfnamefont {B.}~\bibnamefont
  {Hessmo}}, \bibinfo {author} {\bibfnamefont {P.}~\bibnamefont {Usachev}},
  \bibinfo {author} {\bibfnamefont {H.}~\bibnamefont {Heydari}},\ and\ \bibinfo
  {author} {\bibfnamefont {G.}~\bibnamefont {Bj\"ork}},\ }\bibfield  {title}
  {\bibinfo {title} {Experimental demonstration of single photon nonlocality},\
  }\href {https://doi.org/10.1103/PhysRevLett.92.180401} {\bibfield  {journal}
  {\bibinfo  {journal} {Phys. Rev. Lett.}\ }\textbf {\bibinfo {volume} {92}},\
  \bibinfo {pages} {180401} (\bibinfo {year} {2004})}\BibitemShut {NoStop}%
\bibitem [{\citenamefont {van Enk}(2005)}]{Enk05}%
  \BibitemOpen
  \bibfield  {author} {\bibinfo {author} {\bibfnamefont {S.~J.}\ \bibnamefont
  {van Enk}},\ }\bibfield  {title} {\bibinfo {title} {Single-particle
  entanglement},\ }\href {https://doi.org/10.1103/PhysRevA.72.064306}
  {\bibfield  {journal} {\bibinfo  {journal} {Phys. Rev. A}\ }\textbf {\bibinfo
  {volume} {72}},\ \bibinfo {pages} {064306} (\bibinfo {year}
  {2005})}\BibitemShut {NoStop}%
\bibitem [{\citenamefont {Dunningham}\ and\ \citenamefont
  {Vedral}(2007)}]{Dunningham07}%
  \BibitemOpen
  \bibfield  {author} {\bibinfo {author} {\bibfnamefont {J.}~\bibnamefont
  {Dunningham}}\ and\ \bibinfo {author} {\bibfnamefont {V.}~\bibnamefont
  {Vedral}},\ }\bibfield  {title} {\bibinfo {title} {Nonlocality of a single
  particle},\ }\href {https://doi.org/10.1103/PhysRevLett.99.180404} {\bibfield
   {journal} {\bibinfo  {journal} {Phys. Rev. Lett.}\ }\textbf {\bibinfo
  {volume} {99}},\ \bibinfo {pages} {180404} (\bibinfo {year}
  {2007})}\BibitemShut {NoStop}%
\bibitem [{\citenamefont {Heaney}\ \emph {et~al.}(2011)\citenamefont {Heaney},
  \citenamefont {Cabello}, \citenamefont {Santos},\ and\ \citenamefont
  {Vedral}}]{Heaney11}%
  \BibitemOpen
  \bibfield  {author} {\bibinfo {author} {\bibfnamefont {L.}~\bibnamefont
  {Heaney}}, \bibinfo {author} {\bibfnamefont {A.}~\bibnamefont {Cabello}},
  \bibinfo {author} {\bibfnamefont {M.~F.}\ \bibnamefont {Santos}},\ and\
  \bibinfo {author} {\bibfnamefont {V.}~\bibnamefont {Vedral}},\ }\bibfield
  {title} {\bibinfo {title} {Extreme nonlocality with one photon},\ }\href
  {https://doi.org/10.1088/1367-2630/13/5/053054} {\bibfield  {journal}
  {\bibinfo  {journal} {New Journal of Physics}\ }\textbf {\bibinfo {volume}
  {13}},\ \bibinfo {pages} {053054} (\bibinfo {year} {2011})}\BibitemShut
  {NoStop}%
\bibitem [{\citenamefont {Jones}\ and\ \citenamefont
  {Wiseman}(2011)}]{Jones11}%
  \BibitemOpen
  \bibfield  {author} {\bibinfo {author} {\bibfnamefont {S.~J.}\ \bibnamefont
  {Jones}}\ and\ \bibinfo {author} {\bibfnamefont {H.~M.}\ \bibnamefont
  {Wiseman}},\ }\bibfield  {title} {\bibinfo {title} {Nonlocality of a single
  photon: Paths to an einstein-podolsky-rosen-steering experiment},\ }\href
  {https://doi.org/10.1103/PhysRevA.84.012110} {\bibfield  {journal} {\bibinfo
  {journal} {Phys. Rev. A}\ }\textbf {\bibinfo {volume} {84}},\ \bibinfo
  {pages} {012110} (\bibinfo {year} {2011})}\BibitemShut {NoStop}%
\bibitem [{\citenamefont {Brask}\ \emph {et~al.}(2013)\citenamefont {Brask},
  \citenamefont {Chaves},\ and\ \citenamefont {Brunner}}]{Brask13}%
  \BibitemOpen
  \bibfield  {author} {\bibinfo {author} {\bibfnamefont {J.~B.}\ \bibnamefont
  {Brask}}, \bibinfo {author} {\bibfnamefont {R.}~\bibnamefont {Chaves}},\ and\
  \bibinfo {author} {\bibfnamefont {N.}~\bibnamefont {Brunner}},\ }\bibfield
  {title} {\bibinfo {title} {Testing nonlocality of a single photon without a
  shared reference frame},\ }\href {https://doi.org/10.1103/PhysRevA.88.012111}
  {\bibfield  {journal} {\bibinfo  {journal} {Phys. Rev. A}\ }\textbf {\bibinfo
  {volume} {88}},\ \bibinfo {pages} {012111} (\bibinfo {year}
  {2013})}\BibitemShut {NoStop}%
\bibitem [{\citenamefont {Morin}\ \emph {et~al.}(2013)\citenamefont {Morin},
  \citenamefont {Bancal}, \citenamefont {Ho}, \citenamefont {Sekatski},
  \citenamefont {D'Auria}, \citenamefont {Gisin}, \citenamefont {Laurat},\ and\
  \citenamefont {Sangouard}}]{Morin13}%
  \BibitemOpen
  \bibfield  {author} {\bibinfo {author} {\bibfnamefont {O.}~\bibnamefont
  {Morin}}, \bibinfo {author} {\bibfnamefont {J.-D.}\ \bibnamefont {Bancal}},
  \bibinfo {author} {\bibfnamefont {M.}~\bibnamefont {Ho}}, \bibinfo {author}
  {\bibfnamefont {P.}~\bibnamefont {Sekatski}}, \bibinfo {author}
  {\bibfnamefont {V.}~\bibnamefont {D'Auria}}, \bibinfo {author} {\bibfnamefont
  {N.}~\bibnamefont {Gisin}}, \bibinfo {author} {\bibfnamefont
  {J.}~\bibnamefont {Laurat}},\ and\ \bibinfo {author} {\bibfnamefont
  {N.}~\bibnamefont {Sangouard}},\ }\bibfield  {title} {\bibinfo {title}
  {Witnessing trustworthy single-photon entanglement with local homodyne
  measurements},\ }\href {https://doi.org/10.1103/PhysRevLett.110.130401}
  {\bibfield  {journal} {\bibinfo  {journal} {Phys. Rev. Lett.}\ }\textbf
  {\bibinfo {volume} {110}},\ \bibinfo {pages} {130401} (\bibinfo {year}
  {2013})}\BibitemShut {NoStop}%
\bibitem [{\citenamefont {Fuwa}\ \emph {et~al.}(2015)\citenamefont {Fuwa},
  \citenamefont {Takeda}, \citenamefont {Zwierz}, \citenamefont {Wiseman},\
  and\ \citenamefont {Furusawa}}]{Fuwa2015}%
  \BibitemOpen
  \bibfield  {author} {\bibinfo {author} {\bibfnamefont {M.}~\bibnamefont
  {Fuwa}}, \bibinfo {author} {\bibfnamefont {S.}~\bibnamefont {Takeda}},
  \bibinfo {author} {\bibfnamefont {M.}~\bibnamefont {Zwierz}}, \bibinfo
  {author} {\bibfnamefont {H.~M.}\ \bibnamefont {Wiseman}},\ and\ \bibinfo
  {author} {\bibfnamefont {A.}~\bibnamefont {Furusawa}},\ }\bibfield  {title}
  {\bibinfo {title} {Experimental proof of nonlocal wavefunction collapse for a
  single particle using homodyne measurements},\ }\href
  {https://doi.org/10.1038/ncomms7665} {\bibfield  {journal} {\bibinfo
  {journal} {Nature Communications}\ }\textbf {\bibinfo {volume} {6}},\
  \bibinfo {pages} {6665} (\bibinfo {year} {2015})}\BibitemShut {NoStop}%
\bibitem [{\citenamefont {Lee}\ \emph {et~al.}(2017)\citenamefont {Lee},
  \citenamefont {Park}, \citenamefont {Kim},\ and\ \citenamefont
  {Noh}}]{Lee17}%
  \BibitemOpen
  \bibfield  {author} {\bibinfo {author} {\bibfnamefont {S.-Y.}\ \bibnamefont
  {Lee}}, \bibinfo {author} {\bibfnamefont {J.}~\bibnamefont {Park}}, \bibinfo
  {author} {\bibfnamefont {J.}~\bibnamefont {Kim}},\ and\ \bibinfo {author}
  {\bibfnamefont {C.}~\bibnamefont {Noh}},\ }\bibfield  {title} {\bibinfo
  {title} {Single-photon quantum nonlocality: Violation of the
  clauser-horne-shimony-holt inequality using feasible measurement setups},\
  }\href {https://doi.org/10.1103/PhysRevA.95.012134} {\bibfield  {journal}
  {\bibinfo  {journal} {Phys. Rev. A}\ }\textbf {\bibinfo {volume} {95}},\
  \bibinfo {pages} {012134} (\bibinfo {year} {2017})}\BibitemShut {NoStop}%
\bibitem [{\citenamefont {Das}\ \emph {et~al.}(2021{\natexlab{a}})\citenamefont
  {Das}, \citenamefont {Karczewski}, \citenamefont {Mandarino}, \citenamefont
  {Markiewicz}, \citenamefont {Woloncewicz},\ and\ \citenamefont
  {{\.Z}ukowski}}]{1stPaper}%
  \BibitemOpen
  \bibfield  {author} {\bibinfo {author} {\bibfnamefont {T.}~\bibnamefont
  {Das}}, \bibinfo {author} {\bibfnamefont {M.}~\bibnamefont {Karczewski}},
  \bibinfo {author} {\bibfnamefont {A.}~\bibnamefont {Mandarino}}, \bibinfo
  {author} {\bibfnamefont {M.}~\bibnamefont {Markiewicz}}, \bibinfo {author}
  {\bibfnamefont {B.}~\bibnamefont {Woloncewicz}},\ and\ \bibinfo {author}
  {\bibfnamefont {M.}~\bibnamefont {{\.Z}ukowski}},\ }\bibfield  {title}
  {\bibinfo {title} {On detecting violation of local realism with photon-number
  resolving weak-field homodyne measurements},\ }\href@noop {} {\bibfield
  {journal} {\bibinfo  {journal} {arXiv preprint arXiv:2104.10703}\ } (\bibinfo
  {year} {2021}{\natexlab{a}})}\BibitemShut {NoStop}%
\bibitem [{\citenamefont {Das}\ \emph {et~al.}(2021{\natexlab{b}})\citenamefont
  {Das}, \citenamefont {Karczewski}, \citenamefont {Mandarino}, \citenamefont
  {Markiewicz}, \citenamefont {Woloncewicz},\ and\ \citenamefont
  {{\.{Z}}ukowski}}]{2ndPaper}%
  \BibitemOpen
  \bibfield  {author} {\bibinfo {author} {\bibfnamefont {T.}~\bibnamefont
  {Das}}, \bibinfo {author} {\bibfnamefont {M.}~\bibnamefont {Karczewski}},
  \bibinfo {author} {\bibfnamefont {A.}~\bibnamefont {Mandarino}}, \bibinfo
  {author} {\bibfnamefont {M.}~\bibnamefont {Markiewicz}}, \bibinfo {author}
  {\bibfnamefont {B.}~\bibnamefont {Woloncewicz}},\ and\ \bibinfo {author}
  {\bibfnamefont {M.}~\bibnamefont {{\.{Z}}ukowski}},\ }\bibfield  {title}
  {\bibinfo {title} {Can single photon excitation of two spatially separated
  modes lead to a violation of bell inequality via weak-field homodyne
  measurements?},\ }\href {https://doi.org/10.1088/1367-2630/ac0ffe} {\bibfield
   {journal} {\bibinfo  {journal} {New Journal of Physics}\ }\textbf {\bibinfo
  {volume} {23}},\ \bibinfo {pages} {073042} (\bibinfo {year}
  {2021}{\natexlab{b}})}\BibitemShut {NoStop}%
\bibitem [{\citenamefont {Oliver}\ and\ \citenamefont
  {Stroud~Jr}(1989)}]{oliver1989}%
  \BibitemOpen
  \bibfield  {author} {\bibinfo {author} {\bibfnamefont {B.~J.}\ \bibnamefont
  {Oliver}}\ and\ \bibinfo {author} {\bibfnamefont {C.}~\bibnamefont
  {Stroud~Jr}},\ }\bibfield  {title} {\bibinfo {title} {Predictions of
  violations of bell's inequality in an 8-port homodyne detector},\ }\href
  {https://www.sciencedirect.com/science/article/abs/pii/0375960189900364}
  {\bibfield  {journal} {\bibinfo  {journal} {Phys. Lett. A}\ }\textbf
  {\bibinfo {volume} {135}},\ \bibinfo {pages} {407} (\bibinfo {year}
  {1989})}\BibitemShut {NoStop}%
\bibitem [{\citenamefont {Donati}\ \emph {et~al.}(2014)\citenamefont {Donati},
  \citenamefont {Bartley}, \citenamefont {Jin}, \citenamefont {Vidrighin},
  \citenamefont {Datta}, \citenamefont {M.},\ and\ \citenamefont
  {Walmsley}}]{Walmsley14}%
  \BibitemOpen
  \bibfield  {author} {\bibinfo {author} {\bibfnamefont {G.}~\bibnamefont
  {Donati}}, \bibinfo {author} {\bibfnamefont {T.}~\bibnamefont {Bartley}},
  \bibinfo {author} {\bibfnamefont {X.-M.}\ \bibnamefont {Jin}}, \bibinfo
  {author} {\bibfnamefont {M.-D.}\ \bibnamefont {Vidrighin}}, \bibinfo {author}
  {\bibfnamefont {A.}~\bibnamefont {Datta}}, \bibinfo {author} {\bibfnamefont
  {B.}~\bibnamefont {M.}},\ and\ \bibinfo {author} {\bibfnamefont {I.~A.}\
  \bibnamefont {Walmsley}},\ }\bibfield  {title} {\bibinfo {title} {Observing
  optical coherence across fock layers with weak-field homodyne detectors},\
  }\href {https://doi.org/10.1038/ncomms6584} {\bibfield  {journal} {\bibinfo
  {journal} {Nat Commun}\ }\textbf {\bibinfo {volume} {5}},\ \bibinfo {pages}
  {5584} (\bibinfo {year} {2014})}\BibitemShut {NoStop}%
\bibitem [{\citenamefont {Thekkadath}\ \emph {et~al.}(2020)\citenamefont
  {Thekkadath}, \citenamefont {Phillips}, \citenamefont {Bulmer}, \citenamefont
  {Clements}, \citenamefont {Eckstein}, \citenamefont {Bell}, \citenamefont
  {Lugani}, \citenamefont {Wolterink}, \citenamefont {Lita}, \citenamefont
  {Nam}, \citenamefont {Gerrits}, \citenamefont {Wade},\ and\ \citenamefont
  {Walmsley}}]{Walmsley20}%
  \BibitemOpen
  \bibfield  {author} {\bibinfo {author} {\bibfnamefont {G.~S.}\ \bibnamefont
  {Thekkadath}}, \bibinfo {author} {\bibfnamefont {D.~S.}\ \bibnamefont
  {Phillips}}, \bibinfo {author} {\bibfnamefont {J.~F.~F.}\ \bibnamefont
  {Bulmer}}, \bibinfo {author} {\bibfnamefont {W.~R.}\ \bibnamefont
  {Clements}}, \bibinfo {author} {\bibfnamefont {A.}~\bibnamefont {Eckstein}},
  \bibinfo {author} {\bibfnamefont {B.~A.}\ \bibnamefont {Bell}}, \bibinfo
  {author} {\bibfnamefont {J.}~\bibnamefont {Lugani}}, \bibinfo {author}
  {\bibfnamefont {T.~A.~W.}\ \bibnamefont {Wolterink}}, \bibinfo {author}
  {\bibfnamefont {A.}~\bibnamefont {Lita}}, \bibinfo {author} {\bibfnamefont
  {S.~W.}\ \bibnamefont {Nam}}, \bibinfo {author} {\bibfnamefont
  {T.}~\bibnamefont {Gerrits}}, \bibinfo {author} {\bibfnamefont {C.~G.}\
  \bibnamefont {Wade}},\ and\ \bibinfo {author} {\bibfnamefont {I.~A.}\
  \bibnamefont {Walmsley}},\ }\bibfield  {title} {\bibinfo {title} {Tuning
  between photon-number and quadrature measurements with weak-field homodyne
  detection},\ }\href {https://doi.org/10.1103/PhysRevA.101.031801} {\bibfield
  {journal} {\bibinfo  {journal} {Phys. Rev. A}\ }\textbf {\bibinfo {volume}
  {101}},\ \bibinfo {pages} {031801} (\bibinfo {year} {2020})}\BibitemShut
  {NoStop}%
\bibitem [{\citenamefont {Monteiro}\ \emph {et~al.}(2015)\citenamefont
  {Monteiro}, \citenamefont {Vivoli}, \citenamefont {Guerreiro}, \citenamefont
  {Martin}, \citenamefont {Bancal}, \citenamefont {Zbinden}, \citenamefont
  {Thew},\ and\ \citenamefont {Sangouard}}]{Repeater1}%
  \BibitemOpen
  \bibfield  {author} {\bibinfo {author} {\bibfnamefont {F.}~\bibnamefont
  {Monteiro}}, \bibinfo {author} {\bibfnamefont {V.~C.}\ \bibnamefont
  {Vivoli}}, \bibinfo {author} {\bibfnamefont {T.}~\bibnamefont {Guerreiro}},
  \bibinfo {author} {\bibfnamefont {A.}~\bibnamefont {Martin}}, \bibinfo
  {author} {\bibfnamefont {J.-D.}\ \bibnamefont {Bancal}}, \bibinfo {author}
  {\bibfnamefont {H.}~\bibnamefont {Zbinden}}, \bibinfo {author} {\bibfnamefont
  {R.~T.}\ \bibnamefont {Thew}},\ and\ \bibinfo {author} {\bibfnamefont
  {N.}~\bibnamefont {Sangouard}},\ }\bibfield  {title} {\bibinfo {title}
  {Revealing genuine optical-path entanglement},\ }\href
  {https://doi.org/10.1103/PhysRevLett.114.170504} {\bibfield  {journal}
  {\bibinfo  {journal} {Phys. Rev. Lett.}\ }\textbf {\bibinfo {volume} {114}},\
  \bibinfo {pages} {170504} (\bibinfo {year} {2015})}\BibitemShut {NoStop}%
\bibitem [{\citenamefont {Caspar}\ \emph {et~al.}(2020)\citenamefont {Caspar},
  \citenamefont {Verbanis}, \citenamefont {Oudot}, \citenamefont {Maring},
  \citenamefont {Samara}, \citenamefont {Caloz}, \citenamefont {Perrenoud},
  \citenamefont {Sekatski}, \citenamefont {Martin}, \citenamefont {Sangouard},
  \citenamefont {Zbinden},\ and\ \citenamefont {Thew}}]{Repeater2}%
  \BibitemOpen
  \bibfield  {author} {\bibinfo {author} {\bibfnamefont {P.}~\bibnamefont
  {Caspar}}, \bibinfo {author} {\bibfnamefont {E.}~\bibnamefont {Verbanis}},
  \bibinfo {author} {\bibfnamefont {E.}~\bibnamefont {Oudot}}, \bibinfo
  {author} {\bibfnamefont {N.}~\bibnamefont {Maring}}, \bibinfo {author}
  {\bibfnamefont {F.}~\bibnamefont {Samara}}, \bibinfo {author} {\bibfnamefont
  {M.}~\bibnamefont {Caloz}}, \bibinfo {author} {\bibfnamefont
  {M.}~\bibnamefont {Perrenoud}}, \bibinfo {author} {\bibfnamefont
  {P.}~\bibnamefont {Sekatski}}, \bibinfo {author} {\bibfnamefont
  {A.}~\bibnamefont {Martin}}, \bibinfo {author} {\bibfnamefont
  {N.}~\bibnamefont {Sangouard}}, \bibinfo {author} {\bibfnamefont
  {H.}~\bibnamefont {Zbinden}},\ and\ \bibinfo {author} {\bibfnamefont {R.~T.}\
  \bibnamefont {Thew}},\ }\bibfield  {title} {\bibinfo {title} {Heralded
  distribution of single-photon path entanglement},\ }\href
  {https://doi.org/10.1103/PhysRevLett.125.110506} {\bibfield  {journal}
  {\bibinfo  {journal} {Phys. Rev. Lett.}\ }\textbf {\bibinfo {volume} {125}},\
  \bibinfo {pages} {110506} (\bibinfo {year} {2020})}\BibitemShut {NoStop}%
\bibitem [{\citenamefont {Sangouard}\ \emph {et~al.}(2011)\citenamefont
  {Sangouard}, \citenamefont {Simon}, \citenamefont {de~Riedmatten},\ and\
  \citenamefont {Gisin}}]{REV-Repeater}%
  \BibitemOpen
  \bibfield  {author} {\bibinfo {author} {\bibfnamefont {N.}~\bibnamefont
  {Sangouard}}, \bibinfo {author} {\bibfnamefont {C.}~\bibnamefont {Simon}},
  \bibinfo {author} {\bibfnamefont {H.}~\bibnamefont {de~Riedmatten}},\ and\
  \bibinfo {author} {\bibfnamefont {N.}~\bibnamefont {Gisin}},\ }\bibfield
  {title} {\bibinfo {title} {Quantum repeaters based on atomic ensembles and
  linear optics},\ }\href {https://doi.org/10.1103/RevModPhys.83.33} {\bibfield
   {journal} {\bibinfo  {journal} {Rev. Mod. Phys.}\ }\textbf {\bibinfo
  {volume} {83}},\ \bibinfo {pages} {33} (\bibinfo {year} {2011})}\BibitemShut
  {NoStop}%
\bibitem [{\citenamefont {Bachor}\ and\ \citenamefont
  {Ralph}(2019)}]{bachor2019guide}%
  \BibitemOpen
  \bibfield  {author} {\bibinfo {author} {\bibfnamefont {H.}~\bibnamefont
  {Bachor}}\ and\ \bibinfo {author} {\bibfnamefont {T.}~\bibnamefont {Ralph}},\
  }\href {https://books.google.pl/books?id=BMTssgEACAAJ} {\emph {\bibinfo
  {title} {A Guide to Experiments in Quantum Optics}}}\ (\bibinfo  {publisher}
  {Wiley},\ \bibinfo {year} {2019})\BibitemShut {NoStop}%
\bibitem [{\citenamefont {Pan}\ \emph {et~al.}(2012{\natexlab{b}})\citenamefont
  {Pan}, \citenamefont {Chen}, \citenamefont {Lu}, \citenamefont {Weinfurter},
  \citenamefont {Zeilinger},\ and\ \citenamefont {\ifmmode~\dot{Z}\else
  \.{Z}\fi{}ukowski}}]{Pan2012}%
  \BibitemOpen
  \bibfield  {author} {\bibinfo {author} {\bibfnamefont {J.-W.}\ \bibnamefont
  {Pan}}, \bibinfo {author} {\bibfnamefont {Z.-B.}\ \bibnamefont {Chen}},
  \bibinfo {author} {\bibfnamefont {C.-Y.}\ \bibnamefont {Lu}}, \bibinfo
  {author} {\bibfnamefont {H.}~\bibnamefont {Weinfurter}}, \bibinfo {author}
  {\bibfnamefont {A.}~\bibnamefont {Zeilinger}},\ and\ \bibinfo {author}
  {\bibfnamefont {M.}~\bibnamefont {\ifmmode~\dot{Z}\else \.{Z}\fi{}ukowski}},\
  }\bibfield  {title} {\bibinfo {title} {Multiphoton entanglement and
  interferometry},\ }\href {https://doi.org/10.1103/RevModPhys.84.777}
  {\bibfield  {journal} {\bibinfo  {journal} {Rev. Mod. Phys.}\ }\textbf
  {\bibinfo {volume} {84}},\ \bibinfo {pages} {777} (\bibinfo {year}
  {2012}{\natexlab{b}})}\BibitemShut {NoStop}%
\bibitem [{\citenamefont {Clauser}\ and\ \citenamefont {Horne}(1974)}]{CH74}%
  \BibitemOpen
  \bibfield  {author} {\bibinfo {author} {\bibfnamefont {J.~F.}\ \bibnamefont
  {Clauser}}\ and\ \bibinfo {author} {\bibfnamefont {M.~A.}\ \bibnamefont
  {Horne}},\ }\bibfield  {title} {\bibinfo {title} {Experimental consequences
  of objective local theories},\ }\href
  {https://doi.org/10.1103/PhysRevD.10.526} {\bibfield  {journal} {\bibinfo
  {journal} {Phys. Rev. D}\ }\textbf {\bibinfo {volume} {10}},\ \bibinfo
  {pages} {526} (\bibinfo {year} {1974})}\BibitemShut {NoStop}%
\bibitem [{Note1()}]{Note1}%
  \BibitemOpen
  \bibinfo {note} {Like the previous section, there is no violation for the
  vacuum event, i.e., for $n = m = 0$.}\BibitemShut {Stop}%
\bibitem [{\citenamefont {Avesani}\ \emph {et~al.}(2021)\citenamefont
  {Avesani}, \citenamefont {Tebyanian}, \citenamefont {Villoresi},\ and\
  \citenamefont {Vallone}}]{QRNG_appl}%
  \BibitemOpen
  \bibfield  {author} {\bibinfo {author} {\bibfnamefont {M.}~\bibnamefont
  {Avesani}}, \bibinfo {author} {\bibfnamefont {H.}~\bibnamefont {Tebyanian}},
  \bibinfo {author} {\bibfnamefont {P.}~\bibnamefont {Villoresi}},\ and\
  \bibinfo {author} {\bibfnamefont {G.}~\bibnamefont {Vallone}},\ }\bibfield
  {title} {\bibinfo {title} {Semi-device-independent heterodyne-based quantum
  random-number generator},\ }\bibfield  {journal} {\bibinfo  {journal}
  {Physical Review Applied}\ }\textbf {\bibinfo {volume} {15}},\ \href
  {https://doi.org/10.1103/physrevapplied.15.034034}
  {10.1103/physrevapplied.15.034034} (\bibinfo {year} {2021})\BibitemShut
  {NoStop}%
\bibitem [{\citenamefont {Farkas}\ \emph {et~al.}(2021)\citenamefont {Farkas},
  \citenamefont {Guerrero}, \citenamefont {Cariñe}, \citenamefont {Cañas},\
  and\ \citenamefont {Lima}}]{Farkas_2021}%
  \BibitemOpen
  \bibfield  {author} {\bibinfo {author} {\bibfnamefont {M.}~\bibnamefont
  {Farkas}}, \bibinfo {author} {\bibfnamefont {N.}~\bibnamefont {Guerrero}},
  \bibinfo {author} {\bibfnamefont {J.}~\bibnamefont {Cariñe}}, \bibinfo
  {author} {\bibfnamefont {G.}~\bibnamefont {Cañas}},\ and\ \bibinfo {author}
  {\bibfnamefont {G.}~\bibnamefont {Lima}},\ }\bibfield  {title} {\bibinfo
  {title} {Self-testing mutually unbiased bases in higher dimensions with
  space-division multiplexing optical fiber technology},\ }\bibfield  {journal}
  {\bibinfo  {journal} {Physical Review Applied}\ }\textbf {\bibinfo {volume}
  {15}},\ \href {https://doi.org/10.1103/physrevapplied.15.014028}
  {10.1103/physrevapplied.15.014028} (\bibinfo {year} {2021})\BibitemShut
  {NoStop}%
\bibitem [{\citenamefont {Yin}\ \emph {et~al.}(2017)\citenamefont {Yin},
  \citenamefont {Cao}, \citenamefont {Li}, \citenamefont {Liao}, \citenamefont
  {Zhang}, \citenamefont {Ren}, \citenamefont {Cai}, \citenamefont {Liu},
  \citenamefont {Li}, \citenamefont {Dai}, \citenamefont {Li}, \citenamefont
  {Lu}, \citenamefont {Gong}, \citenamefont {Xu}, \citenamefont {Li},
  \citenamefont {Li}, \citenamefont {Yin}, \citenamefont {Jiang}, \citenamefont
  {Li}, \citenamefont {Jia}, \citenamefont {Ren}, \citenamefont {He},
  \citenamefont {Zhou}, \citenamefont {Zhang}, \citenamefont {Wang},
  \citenamefont {Chang}, \citenamefont {Zhu}, \citenamefont {Liu},
  \citenamefont {Chen}, \citenamefont {Lu}, \citenamefont {Shu}, \citenamefont
  {Peng}, \citenamefont {Wang},\ and\ \citenamefont {Pan}}]{Bell-space}%
  \BibitemOpen
  \bibfield  {author} {\bibinfo {author} {\bibfnamefont {J.}~\bibnamefont
  {Yin}}, \bibinfo {author} {\bibfnamefont {Y.}~\bibnamefont {Cao}}, \bibinfo
  {author} {\bibfnamefont {Y.-H.}\ \bibnamefont {Li}}, \bibinfo {author}
  {\bibfnamefont {S.-K.}\ \bibnamefont {Liao}}, \bibinfo {author}
  {\bibfnamefont {L.}~\bibnamefont {Zhang}}, \bibinfo {author} {\bibfnamefont
  {J.-G.}\ \bibnamefont {Ren}}, \bibinfo {author} {\bibfnamefont {W.-Q.}\
  \bibnamefont {Cai}}, \bibinfo {author} {\bibfnamefont {W.-Y.}\ \bibnamefont
  {Liu}}, \bibinfo {author} {\bibfnamefont {B.}~\bibnamefont {Li}}, \bibinfo
  {author} {\bibfnamefont {H.}~\bibnamefont {Dai}}, \bibinfo {author}
  {\bibfnamefont {G.-B.}\ \bibnamefont {Li}}, \bibinfo {author} {\bibfnamefont
  {Q.-M.}\ \bibnamefont {Lu}}, \bibinfo {author} {\bibfnamefont {Y.-H.}\
  \bibnamefont {Gong}}, \bibinfo {author} {\bibfnamefont {Y.}~\bibnamefont
  {Xu}}, \bibinfo {author} {\bibfnamefont {S.-L.}\ \bibnamefont {Li}}, \bibinfo
  {author} {\bibfnamefont {F.-Z.}\ \bibnamefont {Li}}, \bibinfo {author}
  {\bibfnamefont {Y.-Y.}\ \bibnamefont {Yin}}, \bibinfo {author} {\bibfnamefont
  {Z.-Q.}\ \bibnamefont {Jiang}}, \bibinfo {author} {\bibfnamefont
  {M.}~\bibnamefont {Li}}, \bibinfo {author} {\bibfnamefont {J.-J.}\
  \bibnamefont {Jia}}, \bibinfo {author} {\bibfnamefont {G.}~\bibnamefont
  {Ren}}, \bibinfo {author} {\bibfnamefont {D.}~\bibnamefont {He}}, \bibinfo
  {author} {\bibfnamefont {Y.-L.}\ \bibnamefont {Zhou}}, \bibinfo {author}
  {\bibfnamefont {X.-X.}\ \bibnamefont {Zhang}}, \bibinfo {author}
  {\bibfnamefont {N.}~\bibnamefont {Wang}}, \bibinfo {author} {\bibfnamefont
  {X.}~\bibnamefont {Chang}}, \bibinfo {author} {\bibfnamefont {Z.-C.}\
  \bibnamefont {Zhu}}, \bibinfo {author} {\bibfnamefont {N.-L.}\ \bibnamefont
  {Liu}}, \bibinfo {author} {\bibfnamefont {Y.-A.}\ \bibnamefont {Chen}},
  \bibinfo {author} {\bibfnamefont {C.-Y.}\ \bibnamefont {Lu}}, \bibinfo
  {author} {\bibfnamefont {R.}~\bibnamefont {Shu}}, \bibinfo {author}
  {\bibfnamefont {C.-Z.}\ \bibnamefont {Peng}}, \bibinfo {author}
  {\bibfnamefont {J.-Y.}\ \bibnamefont {Wang}},\ and\ \bibinfo {author}
  {\bibfnamefont {J.-W.}\ \bibnamefont {Pan}},\ }\bibfield  {title} {\bibinfo
  {title} {Satellite-based entanglement distribution over 1200 kilometers},\
  }\href {https://doi.org/10.1126/science.aan3211} {\bibfield  {journal}
  {\bibinfo  {journal} {Science}\ }\textbf {\bibinfo {volume} {356}},\ \bibinfo
  {pages} {1140} (\bibinfo {year} {2017})}\BibitemShut {NoStop}%
\bibitem [{\citenamefont {Wengerowsky}\ \emph {et~al.}(2019)\citenamefont
  {Wengerowsky}, \citenamefont {Joshi}, \citenamefont {Steinlechner},
  \citenamefont {Zichi}, \citenamefont {Dobrovolskiy}, \citenamefont {van~der
  Molen}, \citenamefont {Los}, \citenamefont {Zwiller}, \citenamefont
  {Versteegh}, \citenamefont {Mura}, \citenamefont {Calonico}, \citenamefont
  {Inguscio}, \citenamefont {H{\"u}bel}, \citenamefont {Bo}, \citenamefont
  {Scheidl}, \citenamefont {Zeilinger}, \citenamefont {Xuereb},\ and\
  \citenamefont {Ursin}}]{Bell-Submarine}%
  \BibitemOpen
  \bibfield  {author} {\bibinfo {author} {\bibfnamefont {S.}~\bibnamefont
  {Wengerowsky}}, \bibinfo {author} {\bibfnamefont {S.~K.}\ \bibnamefont
  {Joshi}}, \bibinfo {author} {\bibfnamefont {F.}~\bibnamefont {Steinlechner}},
  \bibinfo {author} {\bibfnamefont {J.~R.}\ \bibnamefont {Zichi}}, \bibinfo
  {author} {\bibfnamefont {S.~M.}\ \bibnamefont {Dobrovolskiy}}, \bibinfo
  {author} {\bibfnamefont {R.}~\bibnamefont {van~der Molen}}, \bibinfo {author}
  {\bibfnamefont {J.~W.~N.}\ \bibnamefont {Los}}, \bibinfo {author}
  {\bibfnamefont {V.}~\bibnamefont {Zwiller}}, \bibinfo {author} {\bibfnamefont
  {M.~A.~M.}\ \bibnamefont {Versteegh}}, \bibinfo {author} {\bibfnamefont
  {A.}~\bibnamefont {Mura}}, \bibinfo {author} {\bibfnamefont {D.}~\bibnamefont
  {Calonico}}, \bibinfo {author} {\bibfnamefont {M.}~\bibnamefont {Inguscio}},
  \bibinfo {author} {\bibfnamefont {H.}~\bibnamefont {H{\"u}bel}}, \bibinfo
  {author} {\bibfnamefont {L.}~\bibnamefont {Bo}}, \bibinfo {author}
  {\bibfnamefont {T.}~\bibnamefont {Scheidl}}, \bibinfo {author} {\bibfnamefont
  {A.}~\bibnamefont {Zeilinger}}, \bibinfo {author} {\bibfnamefont
  {A.}~\bibnamefont {Xuereb}},\ and\ \bibinfo {author} {\bibfnamefont
  {R.}~\bibnamefont {Ursin}},\ }\bibfield  {title} {\bibinfo {title}
  {Entanglement distribution over a 96-km-long submarine optical fiber},\
  }\href {https://doi.org/10.1073/pnas.1818752116} {\bibfield  {journal}
  {\bibinfo  {journal} {Proceedings of the National Academy of Sciences}\
  }\textbf {\bibinfo {volume} {116}},\ \bibinfo {pages} {6684} (\bibinfo {year}
  {2019})},\ \Eprint
  {https://arxiv.org/abs/https://www.pnas.org/content/116/14/6684.full.pdf}
  {https://www.pnas.org/content/116/14/6684.full.pdf} \BibitemShut {NoStop}%
\bibitem [{\citenamefont {Grangier}\ \emph {et~al.}(1988)\citenamefont
  {Grangier}, \citenamefont {Potasek},\ and\ \citenamefont {Yurke}}]{GPY}%
  \BibitemOpen
  \bibfield  {author} {\bibinfo {author} {\bibfnamefont {P.}~\bibnamefont
  {Grangier}}, \bibinfo {author} {\bibfnamefont {M.~J.}\ \bibnamefont
  {Potasek}},\ and\ \bibinfo {author} {\bibfnamefont {B.}~\bibnamefont
  {Yurke}},\ }\bibfield  {title} {\bibinfo {title} {Probing the phase coherence
  of parametrically generated photon pairs: A new test of bell's
  inequalities},\ }\href {https://doi.org/10.1103/PhysRevA.38.3132} {\bibfield
  {journal} {\bibinfo  {journal} {Phys. Rev. A}\ }\textbf {\bibinfo {volume}
  {38}},\ \bibinfo {pages} {3132} (\bibinfo {year} {1988})}\BibitemShut
  {NoStop}%
\end{thebibliography}
\end{document}